\documentclass{article}
\pdfoutput = 1
\usepackage[utf8]{inputenc}
\usepackage{jcappub}
\usepackage[dvipsnames]{xcolor}
\usepackage{amssymb,amsmath}
\usepackage{subcaption}
\usepackage{graphicx}

\newcommand{\HI}{\mathrm{HI}}

\author[a]{Shi-Fan Chen,}
\author[a,b]{Emanuele Castorina,}
\author[a,b]{Martin White,}
\author[c]{An\v{z}e Slosar}

\affiliation[a]{Department of Physics, University of California, Berkeley, CA 94720}
\affiliation[b]{Berkeley Center for Cosmological Physics, Berkeley, CA 94720}
\affiliation[c]{Brookhaven National Laboratory, Physics Department, Upton NY 11973}

\emailAdd{shifan\_chen@berkeley.edu}
\emailAdd{ecastorina@berkeley.edu}
\emailAdd{mwhite@berkeley.edu}
\emailAdd{anze@bnl.gov}

\title{Synergies between radio, optical and microwave observations at high redshift}
\date{June 2018}

\abstract{We study synergies between three promising methods to measure $2<z<5$ large-scale structure in the next decade. Optical spectroscopic surveys are the most mature, but become increasingly difficult at $z>2$ and suffer from interloper problems even for spectroscopic surveys. Intensity mapping of the 21-cm signal can cover large volumes with exquisite fidelity, but is limited both by loss of information to foreground cleaning and by lack of knowledge of the mean signal. Cosmic microwave background (CMB) lensing is theoretically very clean, but ultimately measures just the projected variations in density.   We find that cross-correlation between optical and radio can significantly improve the measurement of growth rate. Combining these with the CMB provides a promising avenue to detecting modified gravity at high redshifts, in particular by independently probing the Weyl and Newtonian potentials and by strengthening control of systematics. We find that cross-correlating a Stage {\sc ii} 21-cm survey with DESI quasars with a reasonable brightness temperature prior could enable measurements of the growth rate $f\sigma_8$ at sub 3\% and sub 8\% levels at $z = 3, 4$, representing a factor of 4 and 8 improvement over constraints obtainable from DESI quasars alone. Similarly, cross-correlating 21-cm data with a futuristic LBG survey to $m_{UV}<24.5$ over 1000 square degrees will make possible $f\sigma_8$ measurements at close to 1\% at $z = 3$ and 3\% at $z = 4$, and improve similar constraints at $z = 5$ by close to a factor of 3 to sub-10\% precision. Combining the above with CMB lensing from a Stage 4 CMB survey and LSST data can additionally constrain the gravitational slip $\gamma$ parameter to similar precision at these redshifts, enabling us to test the predictions of general relativity at large scales.
}

\begin{document}
\maketitle
\flushbottom

\section{Introduction}

Intensity mapping with radio interferometers has emerged as a potentially powerful means of efficiently mapping large volumes of the Universe, with sufficient resolution to capture the largest elements of the cosmic web \cite{Kovetz17}.  In this paper we consider specifically 21-cm intensity mapping experiments, which measure the emission signal from neutral hydrogen ($\HI$) in galaxies which trace large-scale structure.  Upcoming experiments, such as CHIME \citep{CHIME}, HIRAX \citep{HIRAX}, BINGO \citep{Bingo} and Tianlai \citep{Tianlai} are designed to measure the clustering of 21-cm emission in redshift space and with high fidelity over the range $0.75<z<2.5$. Other radio observatories not specifically dedicated to cosmological $\HI$ emission, such as SKA \citep{SKACosmo}, are also expected to be sensitive to the 21-cm signal from large-scale structure.  Beyond these, a hypothetical, 21-cm survey focused on $2<z<6$ and large-scale structure was recently proposed and studied in Ref.~\cite{CVDE-21cm} as part of the Cosmic Visions program \cite{Dodelson16}. Such an instrument would yield dramatic advances in our understanding of high $z$ cosmology.

Radio interferometers of the sort envisioned in Ref.~\cite{CVDE-21cm} typically have excellent line-of-sight (radial) resolution but relatively poor angular (transverse) resolution.  For the purposes of measuring large-scale structure this is not a major limitation, as the interferometer resolves the angular scales which can be most easily modeled and which are most correlated with the initial conditions \cite{Pea99,Dod03}.
Of more serious concern are the large, low frequency foregrounds which need to be removed in order to see the cosmological signal and uncertainty in the mean 21-cm brightness temperature ($\bar{T}$), which is proportional to the fractional $\HI$ mass density of the universe and enters as an overall uncertainty in amplitude of cosmological fluctuations in the 21-cm signal.  
One of the primary science goals of many 21-cm intensity mapping surveys, measurements of the cosmic distance scale and the geometry of the Universe using Baryon Acoustic Oscillations (BAO \cite{Weinberg13,PDG18}), is nonetheless possible as BAO experiments resolve the relevant Fourier modes and the uncertainty in $\bar{T}$ does not degrade the measurement of the BAO scale. On the other hand, the uncertainty in $\bar{T}$ presents a serious obstacle to measurements of the growth of structure by means of redshift space distortions (RSD \cite{Kai87,H98}): at large scales (where RSD are well-approximated by linear theory and supercluster/Kaiser infall) the linear growth rate and $\bar{T}$ become perfectly degenerate.  Thus 21-cm data cannot measure the growth rate in the absence of external information on $\bar{T}$.

One possibility for overcoming the inherent limitation imposed by uncertain $\bar{T}$ is to cross-correlate 21-cm surveys with another tracer of the density field at high $z$ \citep{Obuljen18}.  
Cross-correlations will not only be beneficial for the 21-cm science case, but could also help shape future paths for traditional spectroscopy. In addition, these cross correlations will enable independent measurements of $\bar{T}$ at unprecedented precision and constrain models of galaxy evolution at high redshifts as the gas clouds that host most of the neutral hydrogen in the universe are also prime locations for star formation (see e.g.~\cite{McQuinn16}).  Precise RSD measurement requires large volumes to be sampled with large numbers of galaxies, and fulfilling both conditions becomes increasingly challenging at redshift $z\gtrsim2$ in the optical/near infrared band.
In this paper we want to explore the potential of using an optical spectroscopic survey to augment the science return of a future 21-cm experiment observing the high redshift Universe $2<z<6$. 
The future 21-cm facility we have in mind draws heavily from the one discussed in the Cosmic Visions 21-cm white paper \citep{CVDE-21cm}, which was optimized\footnote{The need for a dedicated instrument is set by the desire to make sub-percent distance measurements, which make low-resolution arrays like SKA-MID \citep{Bull16} non optimal for BAO/RSD analyses in the frequency range of interest.} for BAO measurements at $2<z<6$. In the remainder of the paper we will refer to it as the Stage {\sc ii} experiment. More details about Stage {\sc ii} are given in Ref.~\cite{CVDE-21cm}.

To avoid loss of signal due to the foreground subtraction necessary for 21-cm measurements, the optical spectroscopic tracer employed in this scheme must have fine radial resolution.
The first optical instrument we consider is the soon-to-be-commisioned Dark Energy Spectroscopic Instrument (DESI), which will provide a large sample of quasars (QSOs) at intermediate to high redshifts, averaging at least tens of objects per square degree and redshift between $z = 2-4$ over 14,000 deg$^2$ \cite{DESI}. Due to the wide sky area covered and relatively dense numbers, the expected DESI QSOs thus serve as a useful and certain baseline to gauge the synergies of the planned Stage {\sc ii} experiment with optical redshift surveys. At higher redshifts $z \gtrsim 4$ where the cosmological population of QSOs drops precipitously \cite{Kulkarni18}, we explore whether hypothetical, `traditional', Lyman Break Galaxy (LBG) samples \cite{Giavalisco02,Shapley11} can play a similar role, and if so at what limiting magnitude and sky area.

In addition to measuring the growth of large-scale structure, high $z$ surveys offer an almost unique opportunity to test General Relativity (GR) at the largest scales \cite{JaiKho10,Joyce15,Joyce16,Amendola18}.  Particularly interesting is the comparison of the Weyl potential measured by gravitational lensing with the Newtonian potential inferred from the motion of non-relativistic tracers (such as $\HI$ and galaxies).  Within GR, these potentials are equal in the absence of anisotropic stress, but this equality does not hold in many alternative theories of gravity.  Gravitational lensing therefore provides an ideal complement to the 21-cm and galaxy surveys mentioned above, allowing access to the largest volumes of the Universe in a regime, the high redshift Universe, where the predictions are thought to be highly secure.
Weak lensing can be measured by looking at small distortions in the image of source galaxies due to the presence of foreground galaxies, a technique that will be used by surveys like Euclid \citep{EuclidRed} and LSST \citep{LSST}. 
However these imaging surveys will not extend much beyond $z\gtrsim3$, so we will focus on the lensing of the CMB, which probes the intervening matter distribution all the way to the last scattering surface \cite{Seljak96,Lewis06,Hanson10}. 
The highest signal-to-noise, wide-area measurement of CMB lensing currently comes from the {\sl Planck} mission \cite{Planck18-VIII} but in future even more powerful experiments such as the Simons Observatory (SO; \cite{Galitzki18,SO_Science18}) and a Stage 4, ground based CMB experiment (CMB-S4; \cite{CMBS4}) will map large fractions of the sky with ever greater fidelity.
It is worth keeping in mind that, due to foreground cleaning and removal, the cross correlation between the 21-cm signal and any projected statistics (e.g.~lensing) will likely not be observed by any current or future experiment.
Intensity mapping will therefore reduce the degeneracy between astrophysical and cosmological parameters in galaxy-lensing cross correlations through its correlation with the galaxy field.
Another possibility could be to use weak lensing of the 21-cm line itself \cite{Pen04,ZZ06,Pourtsidou15,Foreman18}; this technique is still under development and comes with a lot of technical issues so we will not discuss it further.

We are certainly not the first one to study cross correlation of 21-cm data with optical instruments. 
The first detection of 21-cm line emission was in fact obtained by cross-correlating data collected at the Green Bank Telescope with DEEP2 galaxies \citep{Chang} at $z\simeq0.8$, and many studies have followed since \cite{Switzer13,Masui,Anderson}.
Forecasts for cross correlations at low redshift have been studied in Ref.~\cite{Pourtsidou17} and cross-correlations in N-body and hydrodynamical simulations in \cite{Villaescusa15}.
Our main innovation compared to those papers is that we stress the capability of cross-correlations between 21-cm data and galaxies to enable the calibration of redshift-space distortions measured by 21-cm line intensity mapping, providing growth-rate measurements while marginalizing over the brightness temperature even in the absence of an external prior.

This paper is organized as follows. In Section~\ref{sec:signals}, we describe the two and three dimensional power spectra expected from the 21-cm, spectroscopic galaxy, and CMB lensing data, noting modelling assumptions and potential parameter degeneracies. In Section~\ref{sec:surveys}, we further describe the planned and hypothetical surveys from which the cross-correlations will derive. The forecasting methodology is described in Section~\ref{sec:fisher}. Finally, Sections~\ref{sec:results} and~\ref{sec:cmb} describe our results with and without adding CMB lensing data, respectively. Forecasts for experiments that will exist in the near term, in particular for HIRAX and a DESI-like spectroscopic survey in the Southern hemisphere, are additionally described in Appendix~\ref{sec:HIRAX}. Our conclusions are summarized in Section~\ref{sec:conclusions}.

Throughout this paper we assume a spatially flat $\Lambda$CDM model and for our fiducial parameters we use those derived from the Planck 2018 data plus baryon acoustic oscillations \cite{Planck18-I,Planck18-VI}, specifically $\Omega_m = 0.3106$, $h=0.677$, $n_s=0.96824$ and $\sigma_8=0.811$.

\section{Signals}
\label{sec:signals}

\subsection{The HI signal}

In the post-reionization era most of the hydrogen in the Universe is ionized, and the 21-cm signal comes only from self-shielded, from UV radiation, regions such as galaxies, specifically between the outskirts of disks until where the gas becomes molecular within star-forming regions.
Unfortunately, there are not many observational constraints on the manner in which HI traces galaxies and halos in the high-$z$ Universe, so we are forced to rely on numerical simulations \cite{Dav13,Eagle,VN18} and inferences from other observations \cite{Padmanabhan17,Castorina17}.

On sufficiently large scales, $\HI$ behaves as a linearly biased tracer of the matter field in redshift space.  In this limit the theoretical HI emission power spectrum can be modeled as
\begin{equation}
    P_{\HI}(k,\mu) = \bar{T}^2\left( b_{\HI}+f\mu^2 \right)^2 P_m(k) e^{-(1/2)k^2\mu^2\sigma_{\HI}^2}
    + \bar{T}^2P_{SN}
\label{eqn:PHI}
\end{equation}
where $P_{SN}$ is the shot noise, $f=d\ln D/d\ln a\approx \Omega_m^{0.55}(z)$ is the linear growth rate, $\mu = k_\parallel/k$ is the cosine of the angle to the line-of-sight in Fourier space, $b_{\HI}$ is the (linear) bias of the $\HI$ and $\bar{T}$ is the mean brightness temperature of the $\HI$, defined below.  The finger-of-god term, involving $\sigma_{\HI}$, turns out to have a very small effect for $\HI$ at the scales and redshifts we consider \cite{VN18}. We will set it uniformly to zero in our forecasts.  The important feature to note about Eq.~(\ref{eqn:PHI}) is that the parameters $b_{\HI}$ and $f$ are fully degenerate with $\bar{T}$.  Thus our ability to determine the cosmological parameters is limited by our ability to break this degeneracy, either by additional measurements or an external prior on $\bar{T}$ \cite{Obuljen18}. As external priors can compensate for the effects of additional measurements (in our case a galaxy redshift survey, described below), the strength of our $\bar{T}$ prior will determine the number of additional measurements useful given our prior knowledge, i.e. the amount of data through cross correlations that can give us a sufficient level of further $\bar{T}$ information.

The brightness temperature, $\bar{T}$, is defined in terms of the intensity at frequency $\nu$ as $I_\nu=2k_B\bar{T}(\nu/c)^2=2k_B\bar{T}/\lambda^2$.
The mean cosmological brightness temperature for 21-cm is \cite{Field1959}
\begin{eqnarray}
    \bar{T} &=& \frac{3h_Pc^3\,A_{10}}{32\pi\,m_Hk_B\nu_{21}^2}
    \frac{(1+z)^2}{H(z)}\,\rho_{\HI} \\
    &\simeq& 188h\,(1+z)^2 E(z)\Omega_{\HI}(z)\,{\rm mK}
\end{eqnarray}
with $A_{10}=2.876\times 10^{-15}{\rm s}^{-1}$ the Einstein $A$ coefficient for spontaneous emission, $\nu_{21}\simeq 1.42041\,$GHz the frequency of the emission \cite{Cox2000}, $E(z) = H(z)/H_0$ and the other symbols having their usual meanings. Note that in the first line $h_P$ refers to Planck's constant while in the second line $h$ refers to the Hubble constant in units of $100\,{\rm km}\,{\rm s}^{-1}{\rm Mpc}^{-1}$.  We follow standard convention\footnote{Some previous work (e.g.~Ref.~\cite{Crighton15,Castorina17,VN18}) introduced a related quantity where the density at $z$ is instead normalized by the present day critical density, $\Omega_{\HI,0}\equiv\rho_{\HI}(z)/\rho_{c,0}$.  This differs by a factor of $\rho_c(z)/\rho_c(0)=E^2(z)$ from our definition.} and write the neutral hydrogen density as a fraction of critical at redshift $z$ as $\Omega_{\HI}(z) = \rho_{\HI}(z)/\rho_{c}(z)$.
Unfortunately the value of $\Omega_{\HI}\,h$ is quite uncertain (see e.g.~Refs.~\cite{Padmanabhan15,Crighton15} for recent compilations of data) and it will enter quadratically in the noise power spectrum. 
Existing constraints are obtained by integrating the $\HI$ column density distribution function inferred from QSO spectra. Ref.~\cite{Rao06} measures $10^3\Omega_{\HI}\simeq 0.27\pm 0.10$ at $z\approx 1$, whereas using QSOs in SDSS/BOSS \cite{SDSSDR9,Bird2017} obtained a 5\% measurement at $z\simeq2.3$, $10^3\Omega_{\HI} = 0.086 \pm 0.004$.  Measurements at $z\simeq 2$ are currently modeling limited, but at higher redshift the uncertainties in the determination of the $\HI$ cosmic abundance are dominated by the small number of QSOs spectra publicly available and future surveys will hopefully be able to tighten the constraints.  
We take \cite{Crighton15}
\begin{equation}
    \Omega_{\HI}(z) = 4\times 10^{-4}(1+z)^{0.6}\,E^{-2}(z)
\end{equation}
as our fiducial model, and assume a conservative 10\% prior on $\Omega_{\HI}$, independently of redshift.

To set the $\HI$ bias and `effective' number density of sources we in principle need to know the full distribution of HI within halos. However, given the low angular resolution of intensity mapping surveys each pixel will contain thousands of objects, allowing us to consider only the mean $M_\HI(M_h)$.
Physical intuition and numerical simulations suggest that below a certain halo mass, $M_{\rm min}$, neutral hydrogen will not be self-shielded from UV photons, and that accretion and merging of dark matter halos will likely reduce their HI content over cosmic time. Therefore a well-motivated model of the HI distribution in halos is \cite{Castorina17}
\begin{equation}
    M_{\HI} (M_h;z)  = A(z) M_h^{\alpha(z)} e^{-M_{\rm min}(z)/M_h}
\label{eqn:HI_HOD}
\end{equation}
where we have explicitly written the redshift dependence of the model parameters. 
Within this model we can write \cite{Castorina17,Padmanabhan15,Padmanabhan17}
\begin{equation}
  \rho_c(z)\Omega_{\HI}(z) = \int_0^\infty n(M_h;z) M_{\HI}(M_h;z) \,\mathrm{d}M_h
\end{equation}
and
\begin{equation}
  b_{\HI} = \frac{1}{\rho_c(z) \Omega_{\HI}(z)}\int_0^\infty n(M_h;z) M_{\HI}(M_h;z) b(M_h;z) \,\mathrm{d}M_h
\end{equation}
where $n(M_h;z)$ is the halo mass function and $b(M_h;z)$ the linear halo bias.  We employ the mass and bias functions of Ref.~\cite{SMT2001} throughout.  We have made this particular choice to consistently model the broad redshift range our forecasts cover; however we note that, since the gravitational collapse of the most massive halos tends to be spherical, we expect this choice to somewhat under-predict our biases at the high mass end \cite{CohWhi08,Tinker10}.  The normalization constant $A$ in Eq.~(\ref{eqn:HI_HOD}) can be fixed at each redshift by matching $\Omega_\HI(z)$, but in order to fit for the $M_{\rm min}$ and the slope of the $M_\HI(M_h)$ relation, $\alpha$, we need some information about the clustering of the HI. Lacking such measurements at any redshift we will use the clustering of Damped Lyman-$\alpha$ systems, which contain more than $90\%$ of the neutral hydrogen in the Universe, as a proxy for the HI distribution. In particular BOSS has measured $b_{\rm DLA}= 2.0 \pm 0.1$ at $z=2.3$ \cite{Rafols}.
For the model in Eq.~(\ref{eqn:HI_HOD}), assuming $b_{HI}\simeq b_{DLA}$, the BOSS measurement implies $M_{\rm min} = 5\times 10^9\,h^{-1}M_\odot$ if $\alpha=1$ \cite{Castorina17}.
These numbers agree with an analysis of the most recent hydrodynamical simulations (see Table 6 of Ref.~\cite{VN18}).  Once the model parameters are fixed we can compute the effective number density of HI as 
\begin{equation}
  P_{SN} \equiv \bar{n}_{\rm eff}^{-1} = \frac{1}{[\rho_c(z) \Omega_{\HI}(z)]^2}\int_0^\infty n(M_h;z) M_{\HI}^2(M_h;z)  \,\mathrm{d}M_h \;.
\end{equation}
As the clustering of DLAs is measured only at one redshift with sufficient accuracy, extrapolating the model in Eq.~(\ref{eqn:HI_HOD}) requires some assumptions. If the intensity of the UV background is slowly evolving with redshift we can take both $M_{\rm min}(z)$ and $\alpha(z)$ to be constant. The resulting values of HI clustering and shot noise at $z>2$ are roughly in agreement, at the 20-30\% level, with the hydrodynamical simulations of Ref.~\cite{VN18}. Future data from high redshift QSOs in DESI \cite{DESI} or direct measurement of the 21-cm signal will be required to more accurately model $M_\HI(M_h)$.
Table \ref{tab:obj_props} shows the fiducial value of the bias and effective number density, $\bar{n}_{\rm eff}$, used throughout this paper.

\begin{table}
    \centering
    
    \begin{tabular}{c|cccc|cc|cc|cc|cc}
           &  & & & & &    & \multicolumn{2}{|c|}{$m<24.0$} & \multicolumn{2}{|c|}{$m<24.5$} & \multicolumn{2}{|c}{$m<25.0$} \\
      $z$  & $\chi$ & $\mu_1$ & $\mu_3$ & $k_{\rm nl}$ & $b_{\HI}$ & $\bar{n}_{\rm eff}$ & $\bar{n}_g$ & $b_g$ & $\bar{n}_g$ & $b_g$ & $\bar{n}_g$ & $b_g$  \\ \hline
      2.00 &     3594 &    0.092 &    0.264 &    0.408 & 1.88 & 13.0 & 1.2 &       2.9 &     3.1 &       2.5 &     6.7 &       2.2 \\
      2.50 &     4040 &    0.127 &    0.356 &    0.474 & 2.08 & 20.0 & 0.47 &       3.8 &     1.5 &       3.2 &     3.6 &       2.8 \\
      3.00 &     4404 &    0.167 &    0.449 &    0.540 & 2.30 & 28.0 & 0.16 &       4.8 &     0.60 &       4.1 &     1.7 &       3.6 \\
      3.50 &     4707 &    0.210 &    0.536 &    0.606 & 2.54 & 36.0 & 0.059 &       5.9 &   0.31 &       4.9 &     1.1 &       4.2 \\
      4.00 &     4965 &    0.256 &    0.614 &    0.673 & 2.81 & 44.0 & 0.023 &       6.9 &   0.16 &       5.7 &     0.64 &       4.9 \\
      4.50 &     5187 &    0.303 &    0.681 &    0.740 & 3.11 & 49.0 & 0.012 &       7.8 &   0.084 &       6.5 &    0.35 &       5.6 \\
      5.00 &     5381 &    0.351 &    0.737 &    0.806 & 3.42 & 53.0 & 0.005 &       8.8 &    0.040 &       7.4 &    0.18 &       6.4
    \end{tabular}
    \caption{The properties of the $\HI$ and galaxy samples used in the forecasts.  The comoving distance, $\chi$, is in $h^{-1}$Mpc while the `primary beam' and `pessimistic' wedge parameters, $\mu_1$ and $\mu_3$, are dimensionless.  We estimate the non-linear scale, $k_{\rm nl}$, as the inverse of the rms Zeldovich displacement (in $h{\rm Mpc}^{-1}$).  The $\HI$ bias and `effective' number density (in $10^{-3}\,h^3{\rm Mpc}^{-3}$) assume $M_{\rm min}=5\times 10^9\,h^{-1}M_\odot$ and $\alpha=1$ (see text).  The galaxy bias and number density are for a simple HOD with only central galaxies (see text for discussion).
    }
\label{tab:obj_props}
\end{table}

\subsection{High redshift galaxies and quasars}

As we shall see in later sections, cosmological inferences from 21-cm surveys are limited by the presence of foregrounds and uncertainties in $\bar{T}$.  For this reason we wish to explore how 21-cm surveys can be augmented by classical, high-$z$, galaxy redshift surveys.  We focus here on redshift surveys because the cross-correlation between 21-cm surveys (that lose low $k_\parallel$ modes) and photometric surveys (that contain only low $k_\parallel$ modes) is highly suppressed.
Translating a redshift uncertainty of $\delta z$ into a comoving distance uncertainty of $\delta\chi=[c(1+z)/H(z)]\,[\delta z/(1+z)]$, to probe $k_\parallel=0.03\,h\,{\rm Mpc}^{-1}$ requires $\delta z/(1+z)<0.011-0.015$ at $z=2-5$.  Such photo-$z$ precision is in principle achievable, given enough filters, but primarily at lower redshift and for brighter galaxies \cite{Gorecki14,Laigle16}.  A more reliable approach would be spectroscopic redshifts, if they are available.

Although many spectroscopic tracers of large-scale structure exist (e.g.~the Ly$\alpha$ forest, metal line systems, Ly$\alpha$ emitters, etc.), in this paper we will devote our attention specifically to high redshift quasars (QSOs) or active galactic nuclei and dropout galaxies. Upcoming galaxy surveys such as DESI \cite{DESI} are expected to provide large samples of QSOs out to redshift $z \approx 3$ over wide sky areas, beyond which their population drops precipitously \cite{Kulkarni18}. Alternatively, at high $z$ one can efficiently select samples of galaxies using dropout techniques, e.g.~$u$-band dropouts for $z\sim 3$, $g$-band dropouts for $z\sim 4$ and $r$-band dropouts for $z\sim 5$ \cite{Giavalisco02,Wilson18}.  Magnitude limited dropout samples naturally produce bands in redshift of about $\Delta z\sim 0.5$ with clustering properties that are similar to normal galaxies at $z=0$ \cite{Giavalisco02,Shapley11,Steidel03,Adelberger05,Ouchi05,Lee06,Yoshida06,Reddy08,Hildebrandt09,Vanzella09,Bian13,Bielby13,Barone14,Park16,Malkan17,Ishikawa17,Ono18,Harikane18}.
Spectroscopic follow-up of such samples is challenging, but possible for the brighter objects with existing or planned facilities (see later).

As we deal with large scales we shall approximate the power spectra of these QSO and galaxy populations by \cite{Kai87,H98}
\begin{equation}
    P_g(k) = \left(b_g + f \mu^2\right)^2 P_m(k)e^{-(1/2)k^2\mu^2\sigma^2} + \frac{1}{n_g},
    \label{eq:pkg}
\end{equation}
where the second term accounts for shot noise, assumed Poisson.  For the QSOs the line-of-sight smearing, $\sigma$, will be a combination of redshift errors and fingers of god.  For galaxies the latter dominates if absorption redshifts are used and is still relevant for emission-line redshifts.

We now separately discuss the clustering parameters for QSOs and LBGs. QSOs are highly biased tracers with $b > 3$ at all redshifts of interest; in this paper we will adopt the QSO bias parametrization from eBOSS clustering data, $b_{QSO}(z) = 0.278 ((1 + z)^2 - 6.565) + 2.393$ \cite{Laurent2017}. In addition, the spectra of quasars are very smooth and dominated by a few broad emission features (Lyman-$\alpha$, C\,IV, etc.) that are themselves velocity shifted with respect to the actual redshift of host galaxy. As a result, quasar redshifts are surprisingly poorly understood with typical implied velocity (redshift) errors up to several $100\,{\rm km}\,{\rm s}^{-1}$. In our model these errors are indistinguishable from the finger-of-god effect and we include them as an additional contribution to $\sigma$ of $300 + 400 (z -1.5)\,{\rm km}\,{\rm s}^{-1}$ \cite{eBOSS}.

Since the LBG surveys we discuss in this work are hypothetical, we will model their number density and bias via abundance matching.  Following Refs.~\cite{Sawicki06,VanderBurg10,Ono18} we assume that the the LBG's have close-to-flat continua in the rest frame in the wavelengths of interest and set the K-correction to $2.5\log_{10}(1+z)$.
Taking the UV-luminosity functions from Refs.~\cite{Reddy08}, \cite{Malkan17} and \cite{Ono18} for $z\simeq 2$, $3$ and $4$, we assume galaxies above a given apparent magnitude $M_{UV}$ populate the centers of halos with a soft mass cut from below given by an error function centered at some $M_{\rm min}(M_{UV})$ (see Section 6.2 of Ref.~\cite{Malkan17} for relevant formulae).  Both Refs.~\cite{Malkan17} and \cite{Harikane18} find satellite fractions of less than about $5\%$ for all threshold absolute magnitudes of interest; this trend is reproduced in both semianalytic models \cite{Park16} and HOD fits to clustering on large scales \cite{Jose13,Ishikawa17}, and we thus leave out contributions from satellite galaxies entirely. At each redshift, mass cutoffs are then found by interpolating between the mass cutoffs found as described above at the characteristic redshifts of the above surveys. The number density is then calculated by direct integration up to that mass, and the linear bias similarly computed as the galaxy-number-weighted halo bias.  Number densities and biases as a function of magnitude and redshift are shown in Figure \ref{fig:abundance_matching} and listed in Table \ref{tab:obj_props}.

Finally, we note that our procedure leads to very large bias values, particularly for the brighter, higher redshift galaxies.  This is consistent with observations \cite{Ouchi05,Lee06,Hildebrandt09,Bian13,Bielby13,Park16,Harikane18}.  For such highly biased objects one expects the scale-dependence of the bias to also be significant, and this would need to be addressed in any modeling
(e.g.~\cite{Modi17,Jose17}). Similarly, in modelling both the matter power spectrum and redshift space distortions in linear theory we have necessarily neglected higher-order nonlinear corrections; previous work has suggested that perturbative models can accurately predict the clustering of biased tracers to a significant fraction of the nonlinear scale $k_{\rm nl} = \Sigma^{-1}$ at high $z$ \cite{Carlson13,Vlah16,Foreman16,Modi17}, where $\Sigma$ is mean square one-dimensional displacement in the Zeldovich approximation given by
\begin{equation}
    \Sigma^2 = \frac{1}{6\pi^2} \int P(k) dk.
    \label{eq:sigma}
\end{equation}
For the purposes of forecasts we shall assume that our maximum wavenumber $k_{\rm max}$ is understood to be an effective quantity -- in practice, the higher-order corrections will matter, but the values of those bias parameters will be determined using power-spectrum measurements beyond $k_{\rm max}$ to the accuracy at which they will not affect the precision in measuring quantities of interest. Since development of such analysis goes beyond the purpose of this paper, we leave $k_{\rm max}$ as effectively a free parameter whose value measures our optimism (or pessimism) regarding theoretical model control in future analyses.

\subsection{CMB lensing}

Cosmic microwave background photons are lensed as they traverse the large-scale structure on their journey from the surface of last scattering \cite{Lewis06,Hanson10}.
This lensing effect provides additional sensitivity to low $z$ structure directly from the CMB itself, and is one of the major science drivers for future CMB surveys.

We shall work in terms of the lensing convergence, $\kappa$, and make the Born approximation.  In this limit, $\kappa$ is a projection of the 3D matter density field, weighted by a kernel, $W^\kappa(\chi)$.
Given two projected fields $X,Y$ on the sky the multipole expansion of their angular cross-power spectrum, in the Limber approximation, is
\begin{equation}
  C_\ell^{XY} = \int d\chi\ \frac{W^X(\chi)W^Y(\chi)}{\chi^2}
  \ P_{\delta_{X} \delta_{Y}}\left(k_\perp=\frac{\ell+1/2}{\chi},k_\parallel=0\right)
\label{eqn:ClXY}
\end{equation}
where we have included the lowest order correction to the Limber approximation, $\ell\to\ell+1/2$, to increase the accuracy to $\mathcal{O}(\ell^{-2})$ \cite{LovAfs08,ShaLew08}.  For the lensing convergence $\kappa$ we have $\delta_{\kappa} = \delta_m$, i.e. the matter overdensity, and 
\begin{equation}
  W^{\kappa}(\chi) = \frac{3}{2}\Omega_mH_0^2(1+z)
  \ \frac{\chi(\chi_\star-\chi)}{\chi_\star}
\end{equation}
with $\chi_\star$ the (comoving) distance to last scattering, while for galaxies $\delta_g$ is simply the galaxy overdensity and $W_g \propto dN/dz$. For ease of presentation we have neglected a possible contribution from lensing magnification. To linear order, lensing magnification enters into Equation~\ref{eqn:ClXY} as a redshift-dependent bias for galaxies linear in the slope of the cumulative galaxy distribution function, $n(<m^{\rm th})$, as a function of apparent magnitude. This slope will be well determined for galaxy surveys observing large numbers of objects such as those discussed in this paper; for the sake of simplicity, we have therefore ignored lensing magnification, expecting its effects will enter into any future data analysis parametrized by an essentially-fixed parameter.

To test for possible deviations from general relativity, we also include a factor of
\begin{equation}
    c_{\kappa} = \frac{1+\gamma}{2}, \qquad \gamma = \frac{\Phi}{\Psi}
    \label{eq:gamma}
\end{equation}
for each power of the lensing convergence $\kappa$. Here $\gamma$ is the gravitational slip parameter, equal to unity in general relativity \cite{JaiKho10,Joyce15,Joyce16,Amendola18}, and $c_\kappa$ is the ratio between the Weyl potential and the Newtonian potential $\Psi$. As a hypothetical example, if Poisson's equation for $\Phi$ were modified but $\Psi$ were unaffected, we would have to substitute $\delta_\kappa = c_\kappa \delta_m$ for the lensing potential in Equation~\ref{eqn:ClXY}; this parametrization should, however, also cover any scale-free difference of the relation between matter clustering and the Weyl potential from GR. Following \cite{Schmittfull18}, we parametrize the Newtonian slip as a piecewise function that takes the value $\gamma_i$ in the corresponding redshift bin, treating each $\gamma_i$ as an independent parameter. In a similar spirit many authors have studied the $E_g$ statistic \cite{Zhang07}, a ratio of lensing-galaxy cross correlations and galaxy-velocity cross correlations whose value in our model would be $(1+\gamma)\Omega_{m,0}/2f(z)$. An advantage of our approach is that it can be more directly applied in forward modelling, and the inclusion of complexities such as scale-dependent bias and the dependence on other cosmological parameters (e.g.~$\Omega_m$) can in principle be straightforwardly applied. Our approach also avoids the potential pitfalls of modeling the noise in a data ratio.

\section{Surveys}
\label{sec:surveys}

\subsection{Stage {\sc ii} 21cm experiment}
\label{ssec:stage_ii_exp}

\begin{figure}
    \centering
    \includegraphics[width=\textwidth]{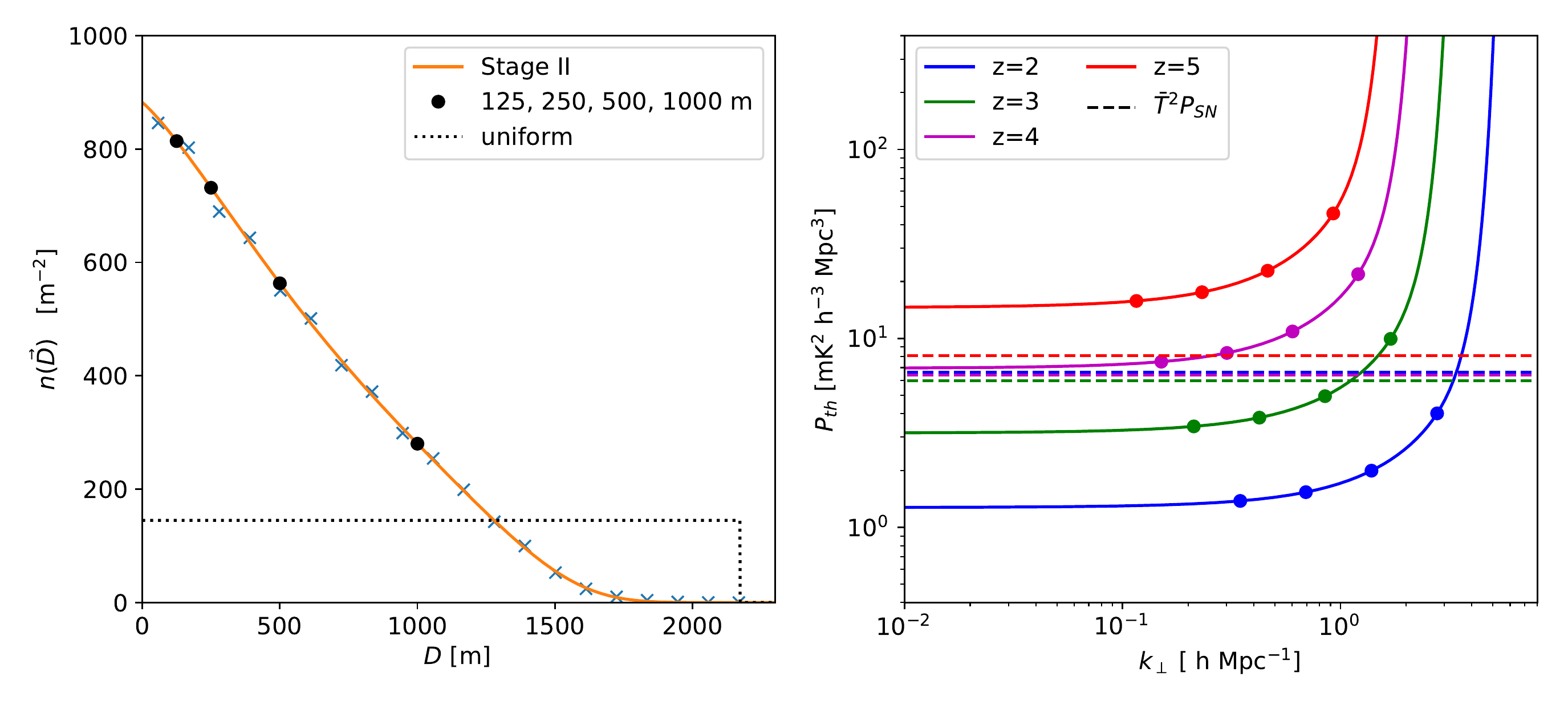}
    \caption{(Left) Baseline distribution for the Stage {\sc ii} experiment in physical units. The distribution $n(\vec{u})$,where $\vec{u} = \vec{D}/\lambda_{\rm obs}$ is the baseline separation in units of observed wavelength, is given by $n(\vec{u}) = \lambda_{\rm obs}^2 n(\vec{D}) $.  The dashed line gives the uniform baseline distribution for the same number of antenna pairs. (Right) Visibility noise spectrum for the baseline distribution of the Stage {\sc ii} experiment run for $t_{obs} = 5 \text{ yrs}$, as a function of wavenumber perpendicular to the line-of-sight, plotted for redshifts $z = 2$, 3, 4, 5. Circular dots on both panels denote baseline separations at 125, 250, 500 and $1000\,$m in increasing order. Dashed lines indicate the shot noise at each corresponding redshift. The thermal noise is dominant over the shot noise at $z = 2,3$ at essentially all scales of interest in this work, and comparable in size to the shot noise at $z = 4, 5$.
    }
    \label{fig:thermal_noise}
\end{figure}

The Stage {\sc ii} 21-cm experiment suggested by Ref.~\cite{CVDE-21cm} consists of a compact, square array of $256\times 256$, fully illuminated $D = 6\,$m dishes observing half the sky with frequencies corresponding to $2<z<6$.  We shall take this as our fiducial survey, though we will also comment upon variations in the dish and array size below.

In an interferometer the fundamental datum is the correlation between two feeds (or antennae) $i$ and $j$, known as a visibility \cite{TMS17}.  On the scales of interest to us, and for an intensity measurement, the visibility measures the Fourier transform of the sky emission times the primary beam at a wavenumber set by the spacing $\vec{u}_{ij}$ of the two feeds in units of the observing wavelength. In particular, such feeds correspond to a comoving wavenumber with component perpendicular to the line-of-sight $k_\perp = 2\pi \vec{u}_{ij}/ \chi(z)$. Thermal fluctuations in the instrument and sky will the introduce additional power into these visibility data, such that the full observed 21-cm signal is given by
\begin{equation}
    P_{21}(k,\mu) = P_{\HI}(k,\mu) + P_{th}.
\end{equation}
The visibility noise is inversely proportional to the number (density) of such baselines, $n(\vec{u})$, normalized such that $\int n(u)d^2u=N_{\rm dish}(N_{\rm dish}-1)/2$.  It is explicitly given by \cite{ZalFurHer04,McQ06,Seo2010,Bull2015,SeoHir16,Wol17,Alonso17,White17,Obuljen18}
\begin{equation}
  P_{th} = T_{\rm sys}^2
    \left( \frac{\lambda^2}{A_{\rm e}} \right)^2
    \left( \frac{4 \pi f_{\rm sky}}{\Omega_p(z)} \right) \frac{1}{n_{\rm pol}\nu_0 t_{\rm obs} n(\vec{u})}
    \frac{d^2V}{d\Omega\,d(\nu/\nu_0)}
    \label{eq:wedge}
\end{equation}
with $\nu_0=1420\,$MHz and $n_{\rm pol} = 2$ the number of polarizations.  The effective area is $A_e = \pi D_e^2/4$, related to the physical area by an aperture efficiency, $\eta_a = 0.7$, such that $D_e^2 = \eta_a D^2$.  We take the field-of-view per pointing to be $\Omega_p(z) = (\lambda / D_e)^2$ -- nearly the square of the FWHM of an Airy disk ($1.028\,\lambda/D$), and approximately the area of a cap of radius $\theta$, $\pi\theta^2$, if we define $\theta=1.22\,\lambda/(2D_e)$.  In a spatially flat model
\begin{equation}
  \frac{d^2V}{d\Omega\,d(\nu/\nu_0)} = \chi^2\,\frac{d\chi}{dz}\,\nu_0\,\frac{dz}{d\nu}
  = \chi^2\,\frac{c\,(1+z)^2}{H(z)} \quad .
\end{equation}
We take $t_{\rm obs}=5 \text{ yrs}$ and the system temperature to be the sum of amplifier noise, sky and ground spill temperatures, $T_{\rm sys} = T_{\rm ampl} + T_{\rm sky} + T_{\rm ground}$, with \cite{CVDE-21cm}
\begin{equation}
    T_{\rm ampl} = 55 \,{\rm K}, \;
    T_{\rm sky} = 2.7\,{\rm K} + 25\,{\rm K}  \left(\frac{\nu_{\rm obs}}{400\,{\rm MHz}} \right)^{-2.75}, \;
    T_{\rm ground} = 30\,{\rm K}   .
\end{equation}
To obtain the baseline distribution, $n(u)$, we have explicitly enumerated all the possible separations between the $256^2\times (256^2-1)/2$ pairs of antennae (Fig.~\ref{fig:thermal_noise}).  For convenience, we fit the distribution with the analytic form
\begin{equation}
    n(\vec{r}) = \lambda_{21}^{-2}(z)
    \ n\left(u=r/\lambda_{21}(z),z\right) = n_0 \frac{c_1 + c_2 x}{1 + c_3 x^{c_4}} e^{-x^{c_5}}
    \quad , \quad x= \frac{r}{N_s D}
\end{equation}
where $r$ is the physical separation between a pair of baselines, $N_s = 256$ is the number of dishes per side\footnote{The same functional form also works for hexagonal arrays with the replacement $c_1=0.5698$, $c_2=-0.5274$, $c_3=0.8358$, $c_4=1.6635$ and $c_5=7.3177$.} of the array, $n_0 = (N_s/D)^2$, and $c_1 = 0.4847$, $c_2 = -0.33$, $c_3 = 1.3157$, $c_4 = 1.5974$ and $c_5=6.8390$ are dimensionless fit parameters, and the redshift dependence is chosen to isolate the redshift-independent distribution of physical distances $r$ (Fig.~\ref{fig:thermal_noise}), such that the distribution $n(\vec{r})$ gives the baseline distribution in physical units. We have checked that this formula is accurate at the percent level for different-sized arrays (down to $N_s = 8$). As shown in Figure ~\ref{fig:thermal_noise}, at $N_s = 256$ the thermal noise levels for the proposed Stage {\sc ii} experiment are smaller than our modelled shot-noise from HI-hosting halos at essentially all scales of interest at $z = 2$ and $3$, and comparable in size to the shot noise up to the maximum redshift $z = 5$ that we consider. That the total noise is dominated by shot noise at lower redshifts and the thermal noise is comparable to shot noise at higher redshifts is in contrast with previous surveys, which were thermal-noise dominated \cite{Cohn16,Obuljen18,CHIME,HIRAX,Tianlai}.

\begin{figure}
    \centering
    \includegraphics[width=\textwidth]{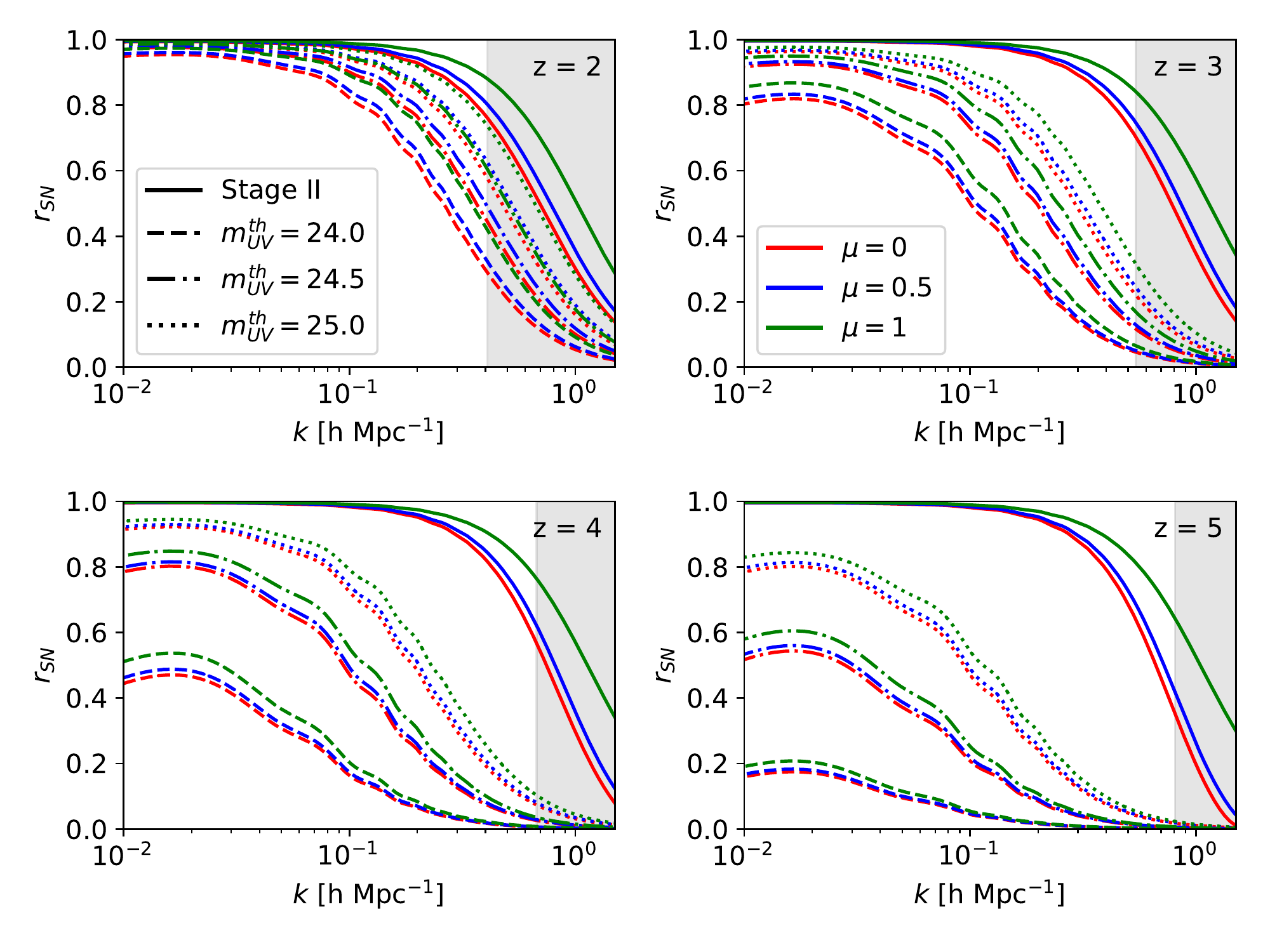}
    \caption{The theory power as a fraction of total power, including noise, for both 21-cm measurements from the proposed Stage {\sc ii} experiment (Section~\ref{ssec:stage_ii_exp}) and hypothetical high-redshift galaxy surveys at limiting magnitudes of $m_{UV}^{\rm th} = 24$, $24.5$ and  $25$, plotted against three-dimensional wavenumber $k$ at redshifts $z = 2$, 3, 4 and 5 for three values of $\mu = k_\parallel/k$. The region to the right of the nonlinear scale $k_{\rm nl}$, beyond which we expect modeling to break down, is shaded in gray. The 21-cm signal, described in Eq.~\ref{eqn:PHI}, is expected to be signal-dominated at all scales of interest. In comparison, the galaxy power spectrum will be noise dominated at most modes we model except at the lowest redshifts.}
    \label{fig:snr}
\end{figure}

\begin{figure}
    \centering
    \includegraphics[width=\textwidth]
    {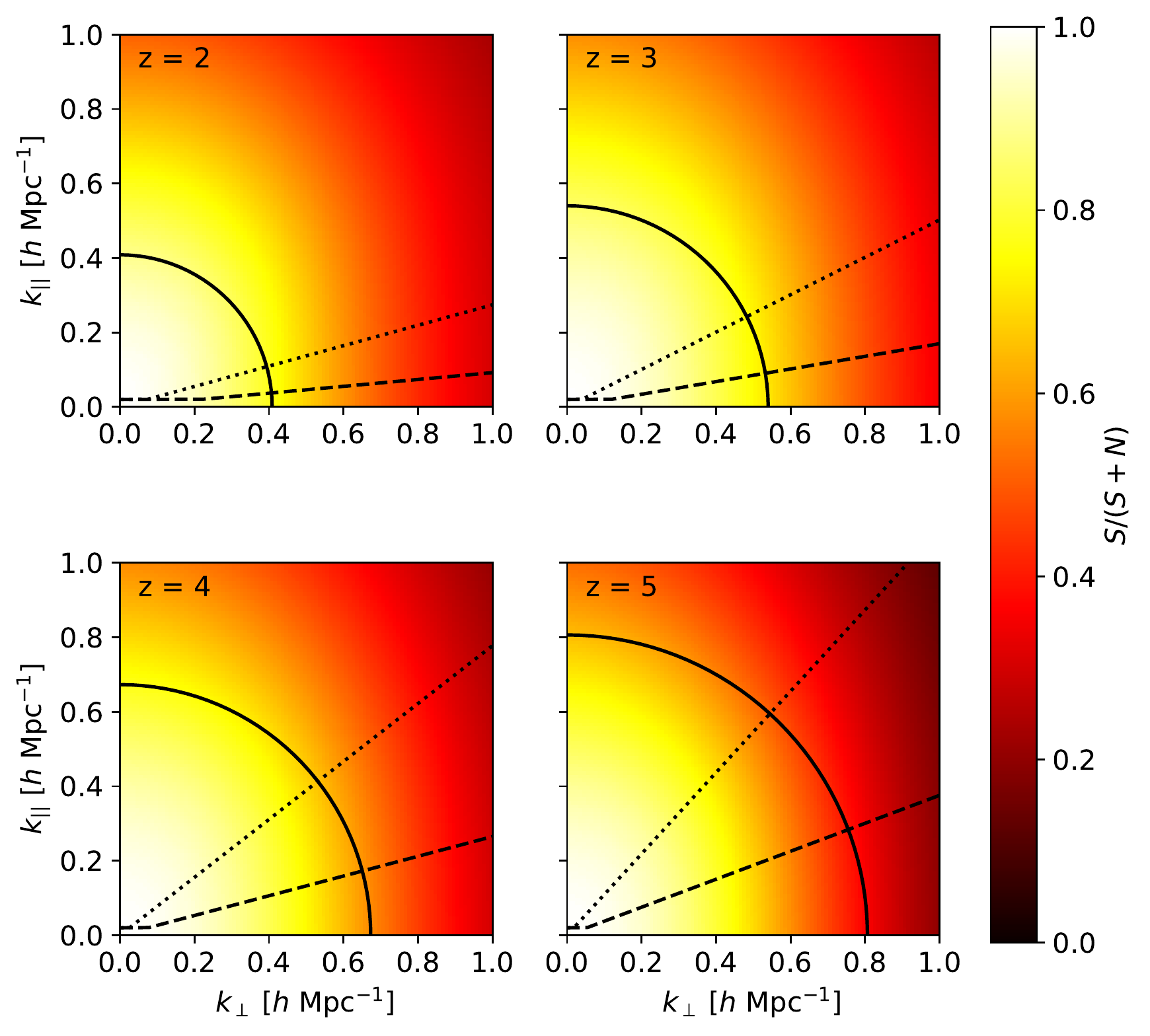}
    \caption{The expected $\HI$ signal as a fraction of total observed power in the $k_\perp-k_\parallel$ plane at redshifts $z = 2$, 3, 4, $5$.  The dashed and dotted lines indicated the angular $\mu$ cut imposed by the foreground wedge in our ``optimistic'' and ``pessimistic'' scenarios respectively (see text). At sufficiently low $k_\perp$ both these wedges intersect with the minimum $k_\parallel$ cut imposed by foregrounds. We have not attempted to model the effects of foregrounds but rather assume that modes inside the wedge will have to be discarded (see text). The solid black curve indicates the nonlinear scale, $k_{\rm nl}$, at each redshift.}
    \label{fig:kplane}
\end{figure}

Figure~\ref{fig:snr} shows the fraction $r_{i, SN}$ of total power taken up by the signal, i.e.
\begin{equation}
    r_{i, SN}(k,\mu,z) = \frac{P_{i} - P_{i,N}}{P_{i}},
\end{equation}
for both the expected 21-cm signal and galaxy power spectra (described below) at $z=2-5$. Here $P_{i}$ is the total power of the i$^{th}$ species ($i = 21, g$) and $P_{i,N}$ is the power from noise alone, where we assume for galaxies $P_{g,N}$ is given by the shot noise. While the cosmological signal from galaxies drops sharply at high redshifts and small scales, the 21-cm signal from the Stage {\sc ii} experiment would be signal-dominated at all redshifts of interest up to the nonlinear scale where modelling breaks down, suggesting potential for significant gains over galaxies alone if the $\bar{T}$ degeneracy can be broken by pusing to high wavenumbers.

Despite the above, the major difficulty facing upcoming 21-cm experiments is astrophysical foregrounds, primarily free-free and synchrotron emission from the Galaxy and unresolved point sources \cite{Furlanetto06,Shaw14,Pober15,Seo16,Cohn16}.
The precise range of scales accessible to 21-cm experiments after foreground removal is currently a source of debate.  We do not attempt to model foreground subtraction explicitly, but take into account its effects by restricting the range of the $k_\parallel - k_{\perp}$ plane we use (see Fig.~\ref{fig:kplane}).  There are two regions of this plane we could lose to foreground removal.  The first is low $k_\parallel$ modes, i.e.~modes close to transverse to the line-of-sight, which are swamped by the (large amplitude but spectrally smooth) foregrounds. We follow \cite{Shaw14,Shaw15} and assume all modes with $k_\parallel < 0.02 \, h \text{Mpc}^{-1}$ are unusable. We have checked that our forecasts are relatively insensitive to this cut -- higher $k_\parallel$ cuts (such as the $0.1 \, h \text{Mpc}^{-1}$ limit suggested in \cite{Pober15}) do not substantively affect our results.  In addition, non-idealities in the instrument lead to leakage of foreground information into higher $k_\parallel$ modes.  This is usually phrased in terms of a foreground ``wedge'' which renders modes with $k_\parallel/k_\perp<\sin\theta_w\,\mathcal{R}$ unusable, with \cite{Datta10,Morales12,Parsons12,Shaw14,Liu14,Pober15,Seo16,Cohn16}
\begin{equation}
    \mathcal{R} \equiv \frac{\chi(z)\,H(z)}{c(1+z)}
    = \frac{E(z)}{1+z}\int_0^z\frac{dz'}{E(z')} \quad .
    \label{eqn:wedgeR}
\end{equation}
The last equality assumes a spatially flat Universe.
This corresponds to a cut $\mu<X/\sqrt{1+X^2}$ with $X=\sin\theta_w\,\mathcal{R}$, which is listed in Table~\ref{tab:obj_props} as a function of $z$.  Information in this wedge is not irretrievably lost, but it becomes more and more difficult to extract the deeper into the wedge one pushes. Our fiducial choice will be the `primary beam' wedge defined with $\theta_w\approx 1.22\,\lambda/2D_e$, where the factor of two in the denominator gives an approximate conversion between the first null of the Airy disk and its FWHM as described above.  We shall contrast this ``optimistic'' assumption with the ``pessimistic'' case $\theta_w\approx 3\times 1.22\lambda/2D_e$.  Figure~\ref{fig:kplane} shows the two wedge scenarios at four redshifts, $z = 2$, 3, 4 and $5$, in the $k_\perp-k_\parallel$ plane. At each redshift the pessimistic scenarios removes a significantly larger number of modes than the optimistic scenario. Note that we have not attempted to model the effects of foregrounds within the wedge in the figure, since for our purposes the modes enclosed will simply be discarded -- the interested reader is pointed to Figures 1-6 in ref.~\cite{Morales12} for a clear pedagogical exposition of why the wedge arises, and to refs.~\cite{Byrne19,Pober15} for plots of the simulated and observed foreground footprints in the wedge. We also note that the wedge could potentially be even larger, possibly up to $k_\parallel/k_\perp<\mathcal{R}$, sometimes known as the ``horizon wedge''.  We do not consider this possibility, as it renders the 21-cm experiment ineffective as a large-scale structure probe.

\subsection{DESI}

As part of its Ly$\alpha$ forest survey, DESI in the near term will provide a large sample of QSOs at intermediate to high redshifts ($z = 2-4$). For our forecasts we assume the DESI QSOs follow a top hat distribution in each redshift bin with the amplitude of $dN/dz$ interpolated from Table 2.7 of Ref.~\cite{DESI} over a sky area of 14,000 square degrees. Specifically, DESI will yield a comoving density of $\sim 2 \times 10^{-5} \; h^3 \text{Mpc}^{-3}$ at $z = 2$, but fall dramatically to $\sim 8 \times \; 10^{-7} h^3 \text{Mpc}^{-3}$ by $z = 4$. These QSOs are highly biased tracers, with bias values rising from about $3.1$ to $7.5$ over the same redshift range.

In the lower redshift universe, DESI will observe a large sample of Emission Line Galaxies (ELGs) between $z = 0.65-1.65$, with projected densities of between one and two thousand per redshift per square degree below $z = 1.25$. These galaxies fall into the redshift range of more near-term 21-cm experiments such as CHIME and HIRAX \cite{CHIME,HIRAX}.   Forecasts combining these surveys are discussed briefly in Appendix \ref{sec:HIRAX}. In practice, galaxy densities are sufficiently high at lower redshifts that adding 21-cm data  does not offer significant improvement.

\subsection{Dropout surveys}

An exciting opportunity for selecting high-$z$ galaxies is through dropout techniques\footnote{Some of the dropouts will naturally be strong line emitters, including Ly$\alpha$ emitters \cite{Cooke09}.  LAEs could also be efficiently selected at specific redshifts through narrow-band imaging.  For a narrow enough selection a non-zero cross-correlation can be achieved without spectroscopy, but in general follow-up spectroscopy is necessary.  This approach can become expensive due to the narrow redshift range obtained for each filter.}, followed by optical and near-IR spectroscopy to obtain redshifts.  We choose not to model the dropout surveys in any detail, but rather ask what sky area and number density would be required to achieve the desired science.  This then sets the `expense' of the required survey, using existing or planned spectroscopic facilities
(e.g.~DESI \cite{DESI}, 4MOST \cite{Roelof16}, Subaru PFS \cite{Takada14}, Keck-FOBOS\footnote{https://github.com/Keck-FOBOS}, or a future ESO spectroscopic telescope \cite{Pasquini16}).

\begin{figure}
    \centering
    \includegraphics[width=\textwidth]{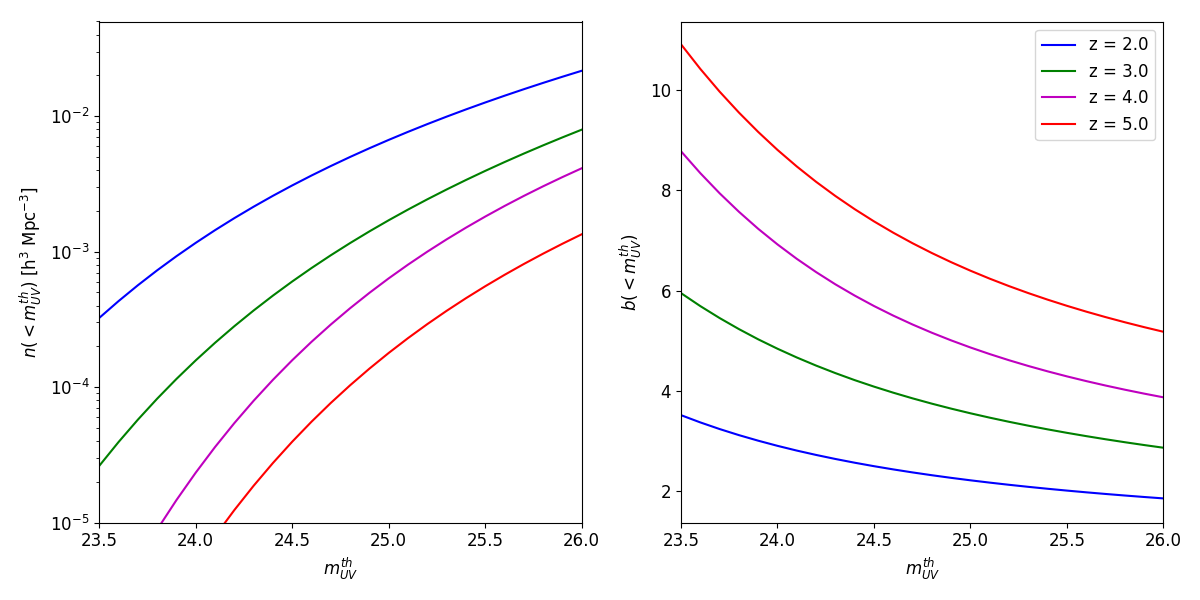}
    \caption{Abundance matched number densities and biases for galaxies above a given apparent magnitude $m_{UV}^{th}$ at redshifts $z = 2, 3, 4, 5.$ Luminosity functions are taken from Refs.~\cite{Reddy08}, ~\cite{Malkan17} and \cite{Ono18}, with corresponding mass cuts, implemented as a soft error function turn-off, calculated at the characteristic redshifts of each survey redshift bin. We employ the mass and bias functions of \cite{SMT2001} throughout. }
    \label{fig:abundance_matching}
\end{figure}

In what follows, we assume a tophat $dN/dz$ for the dropout sample with FWHM $\Delta z=0.5$, centered on the fiducial redshift of the sample.  This is only an approximation to a dropout selection (see e.g.~Fig.~2 of Ref.~\cite{Ishikawa17} or Fig.~6 of Ref.~\cite{Ono18}), but it is sufficient for our purposes.  In addition, a major problem for dropout samples is the presence of interlopers \cite{Wilson18}. We shall assume that a fraction, $f_{\rm int}$, of the imaging sample are interlopers, i.e.~they are not LBGs in the desired redshift range (an $f_{\rm int}<1$ can also account for galaxies for which redshifts are hard to obtain \cite{Cooke09}). The observed power spectrum will thus actually be
\begin{equation}
    P_{g,obs} = (1 - f_{\rm int})^2 P^{\rm th}_g + f_{\rm int}^2 P^{\rm th}_{\rm int} + \frac{1}{n_{g,{\rm obs}}},
    \label{eqn:interlopers}
\end{equation}
where $P^{\rm th}_g$ and $P^{\rm th}_{\rm int}$ are the theoretical predictions for the dropout galaxy and interloper power spectra, respectively. Note that the shot noise, given by the inverse of the total observed number density, is unchanged from the interloper free case. The interloper fraction $f_{\rm int}$ is in principle a free parameter in our modelling and would need to be accounted for in our analysis; indeed, in the event that the interlopers have zero power it is perfectly degenerate with the amplitude of $(b_g + f\mu^2)$ and will hinder our ability to break the $b_{\HI}-\bar{T}$ degeneracy via cross correlation. For our forecasts, we take a fixed, fiducial value of $f_{\rm int}=0.5$ and assume future independent analyses will provide a strong enough prior on $f_{\rm int}$ that we can treat it as a fixed parameter.

\subsection{CMB surveys}

The study of CMB lensing is entering an area of rapid experimental advance, with the Simons Observatory (SO) \cite{SO_Science18} and CMB-S4 \cite{CMBS4} promising to map large fractions of the sky with much higher S/N than currently achievable.  We shall follow standard practice and assume that the lensing noise is dominated by fluctuations in the primary CMB and detector noise; the full expression is given in e.g.~Ref.~\cite{Hanson10}.
In this paper we will focus on synergies with CMB-S4, assuming a $1.4^\prime$ beam and $1\,\mu$K-arcmin map noise.  Following \cite{Schmittfull18}, we divide the $EB$ noise by a factor of 2.5 to approximate expected improvements from iterative lens reconstruction \cite{CMBS4}. We find our forecasts are relatively insensitive to the assumed angular resolution, but are a stronger function of the map noise level.

\subsection{LSST}

Other authors (e.g. Refs~\cite{Vallinotto12,Pearson14,Takeuchi14, Modi17,Banerjee18,Mishra-Sharma18, Schmittfull18,Yu18} have noted significant gains achievable by combining CMB lensing data with upcoming photometric surveys such as LSST. Since the lensing deflection integrates the cumulative effect of gravitational potentials up to the surface of last scattering, cross-correlating with such a galaxy sample  allows us to `pick out'' data at specific redshift intervals. To this end we adopt the ``optimistic'' LSST number densities from taken from \cite{Gorecki14}, assuming a magnitude of $i < 27$ with signal-to-noise cutoff at $SNR = 5$ after three years of observation.  Due to the higher redshift uncertainties associated with photometric surveys we analyze only the two dimensional angular power spectra, $C_\ell$, of these galaxies. To maintain consistency with \cite{Yu18, Schmittfull18}, we have calculated these using the above $dN/dz$ as well as the redshift-dependent linear bias parameter $b(z)=1+z$ \cite{LSST}.  This scaling of the bias is very roughly consistent with measurements of faint, high-$z$ galaxy clustering but could be improved in future forecasts. Unless otherwise mentioned, all CMB analyses in this paper will include LSST data by default where available.

\section{Forecasting methodology}
\label{sec:fisher}

We investigate the constraining power of future surveys through a Fisher matrix formalism \citep{Tegmark,White09}. We work at the level of the fields and combine our measurements into a covariance matrix
\begin{equation}
    C(k) = 
    \begin{pmatrix}
        P_{21}(k)   & P_{21,g}(k) \\
        P_{21,g}(k) & P_{g}(k)
    \end{pmatrix}
\end{equation}
which is diagonal in wavenumber. The auto-power spectra are as defined above; we take the cross-power spectrum to be the geometric mean of the two theory power spectra, and assume the shot noise is negligible. Confounding observational effects need also to be included in the spectra in $C(k)$; for example, should interlopers exist their effect on the data would have to be integrated into our forecasts by swapping $P_g$ above for $P_{g,obs}$ as defined in Equation~\ref{eqn:interlopers}.  We do not, however, include the Alcock-Paczynski effect \cite{Alcock1979}. We can then construct a Fisher matrix for the parameters $\theta_i = \{ \bar{T}, f\sigma_8, b_{HI}\sigma_8, b_g\sigma_8\}$, given by
\begin{equation}
    F_{ij} = V_{\rm obs} \int \frac{d^3k}{(2\pi)^3}\ \frac{1}{2} \text{Tr}
    \left[ C^{-1} \frac{\partial C}{\partial \theta_i}
           C^{-1} \frac{\partial C}{\partial \theta_j} \right]
\end{equation}
where $i$, $j$, run over the parameters and $V_{\rm obs}$ is overlapping comoving volume of the galaxy and interferometric surveys.
The latter can be written as $V_{\rm obs} = f_{\rm sky} V(z_1,z_2),$ i.e.~the product of the overlapping sky fraction and the total comoving volume inside the redshift limits $z_1$, $z_2$ of the appropriate redshift bin of the galaxy survey. Fisher matrices constructed using only individual tracers are computed by using covariance matrices consisting of only the autocorrelation.

An additional specification is the wavenumber ranges over which we integrate the Fisher matrices. The volume of the survey sets a fundamental limit on the minimum wavenumber, or maximum scale, that we can probe, which we take to be $k_{\rm min} = 2 \pi / (V_{\rm obs})^{1/3}$.
As our fiducial choice we shall assume an upper limit $k_{\rm max} = 0.4 \, h\, \text{Mpc}^{-1}$, independent of redshift.  This is larger than the value often assumed for forecasts at lower redshift, because the field is expected to be more linear at early times.  We shall discuss the impact of varying $k_{\rm max}$ with redshift later.
Finally, as described above (see Eq.~\ref{eq:wedge} and surrounding text) the particularities of the 21-cm signal impose additional integration limits: foreground removal sets a lower bound on $k_\parallel$ while the wedge provides a lower bound on $\mu$ for usable $k$-modes. In what follows we will generally assume the optimistic primary wedge, but will explore the effects of the more restrictive ``pessimistic'' wedge as well.

The above Fisher matrix contains information only from the overlapping area of the galaxy and interferometric surveys. Since in general the interferometric survey will span a significantly larger area than the galaxy survey the non-overlapping area must also be considered. To a good approximation, the 21-cm signal from this area will be uncorrelated with that of the overlap, so it is sufficient to add a Fisher matrix computed using only the HI-HI autocorrelation (Appendix~\ref{sec:overlap}). Similarly, as the full covariance between galaxies and the 21-cm signal can be obtained with significance only above the wedge, to include the available information at $\mu < \mu_{\rm min}$ we must add on the galaxy-autocorrelation-derived Fisher matrices from this region in $k$-space. This is equivalent to doing the full $k$-space integral in the case where 21-cm noise goes to infinity in the wedge.

To include constraints from CMB lensing we incorporate Fisher matrices derived from the angular covariances
\begin{equation}
    C_{2D}(\vec{l}) = 
                    \begin{pmatrix}
                        C^{\kappa \kappa}_l   & C^{\kappa g}_l \\
                        C^{\kappa g}_l & C^{gg}_l.
    \end{pmatrix}
\end{equation}
In the above we have not included possible correlations with 21-cm data; as seen in Equation~\ref{eqn:ClXY}, the above angular covariances probe Fourier modes with $k_\parallel \sim 0$, where foreground subtraction would render 21-cm data useless. Indeed, a galaxy distribution with our assumed width $\Delta z = 0.5$ would smooth out modes above $k_\parallel \simeq 6 \times 10^{-3}\,h\, {\rm Mpc}^{-1}$, substantially below even more generous assumptions about astrophysical foreground removal, not to mention the lensing kernel. This is equivalent to computing the corresponding three-dimensional Fisher matrix and summing over only those wave modes with $k_\parallel \lesssim 2 \pi / \Delta \chi.$ 

Finally, the information from the surveys can be combined with parameter constraints from independent data to further increase our constraining power. This is implemented by adding the Fisher matrices from these data to the above. In particular, in light of existing measurements on $\Omega_{\HI}$ discussed above, we use a fiducial prior of 10\% on the brightness temperature $\bar{T}$.

\section{Results}
\label{sec:results}

\subsection{Synergies with DESI}

\begin{figure}
    \centering
    \includegraphics[width=\textwidth]{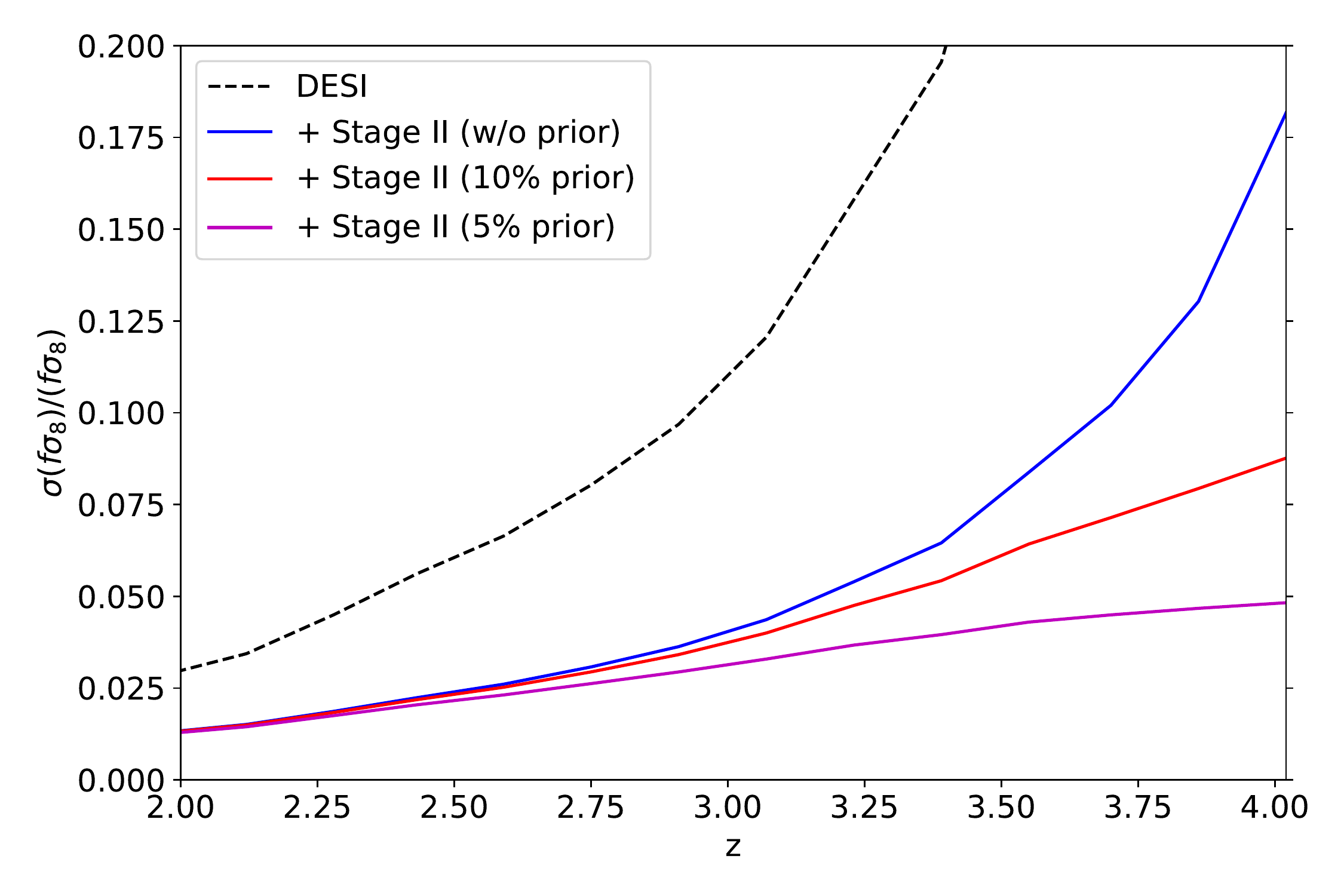}
    \caption{Constraints on $f\sigma_8$ using DESI and 21-cm data from the hypothetical Stage {\sc ii} experiment, plotted for 10\% and 5\% brightness-temperature $\bar{T}$ priors, as well as no prior. Adding 21-cm data improves constraints from DESI by a factor of two at $z = 2$ and by more than a factor of four at $z = 4$. At the highest redshifts probed, the joint constraint would however be dominated by our (conservative) 10\% prior on the brightness temperature, $\bar{T}$.}
    \label{fig:desi_constraints}
\end{figure}

We begin by considering the extent to which RSD measurements of the DESI QSO sample at $z = 2-4$ can be improved by cross-correlation with 21-cm data from the hypothetical Stage {\sc ii} experiment. Figure~\ref{fig:desi_constraints} shows the constraints on $f \sigma_8$ that can be obtained by combining both datasets compared to those from DESI QSOs alone, assuming full survey overlap and redshift bin of $\Delta z = 0.5$. Adding the 21-cm data, shown with the blue line, improves the RSD measurement by around a factor of 2 at $z = 2$, giving an $f \sigma_8$ constraint close to a percent.  The level of improvement increases to higher redshift and by $z=4$ is a factor of 4.  In particular, the $>10\%$ constraint at redshift $z = 3$ from DESI alone becomes a $<5\%$ constraint with the addition of 21-cm data.

A qualitative shift occurs for $f \sigma_8$ constraints at the highest redshifts probed by DESI. At $z = 4$, the addition of a 10\% prior on the brightness temperature $\bar{T}$ leads to a factor of 2 improvement, resulting in a sub-ten percent constraint, and a 5\% prior would further improve the constraint to the level of the prior;  in contrast, the $f \sigma_8$ constraints at $z = 2$ are essentially unaffected by these priors. This indicates that such priors would dominate over the galaxy data in breaking the $f\sigma_8-\bar{T}$ degeneracy. This is unsurprising: the DESI QSO population declines precipitously (by more than an order of magnitude) from $z=2$ to $4$.
We thus need a spectroscopic galaxy sample that is denser than the DESI QSOs at high redshift to qualitatively improve our RSD constraints past what can be obtained from independent $\bar{T}$ priors and 21-cm data alone.

\subsection{Synergies with a future LBG Survey}

In this subsection we explore potential synergies in cross-correlating the Stage {\sc ii} 21-cm experiment with a spectroscopic LBG survey at redshifts $z = 3-5$ where the cross-correlations with DESI become $\bar{T}$ prior dominated. Since such a survey is hypothetical, we thus explore the relative benefits of various survey configurations, primarily varying sky coverage and depth, in three redshift bins $z = 3$, 4 and $5$. For our fiducial forecasts we assume no finger-of-god or undetermined interloper fractions, and a brightness temperature $\bar{T}$ prior of 10\%.  We revisit each assumption later.

\begin{figure}
    \centering
    \includegraphics[width=\textwidth]{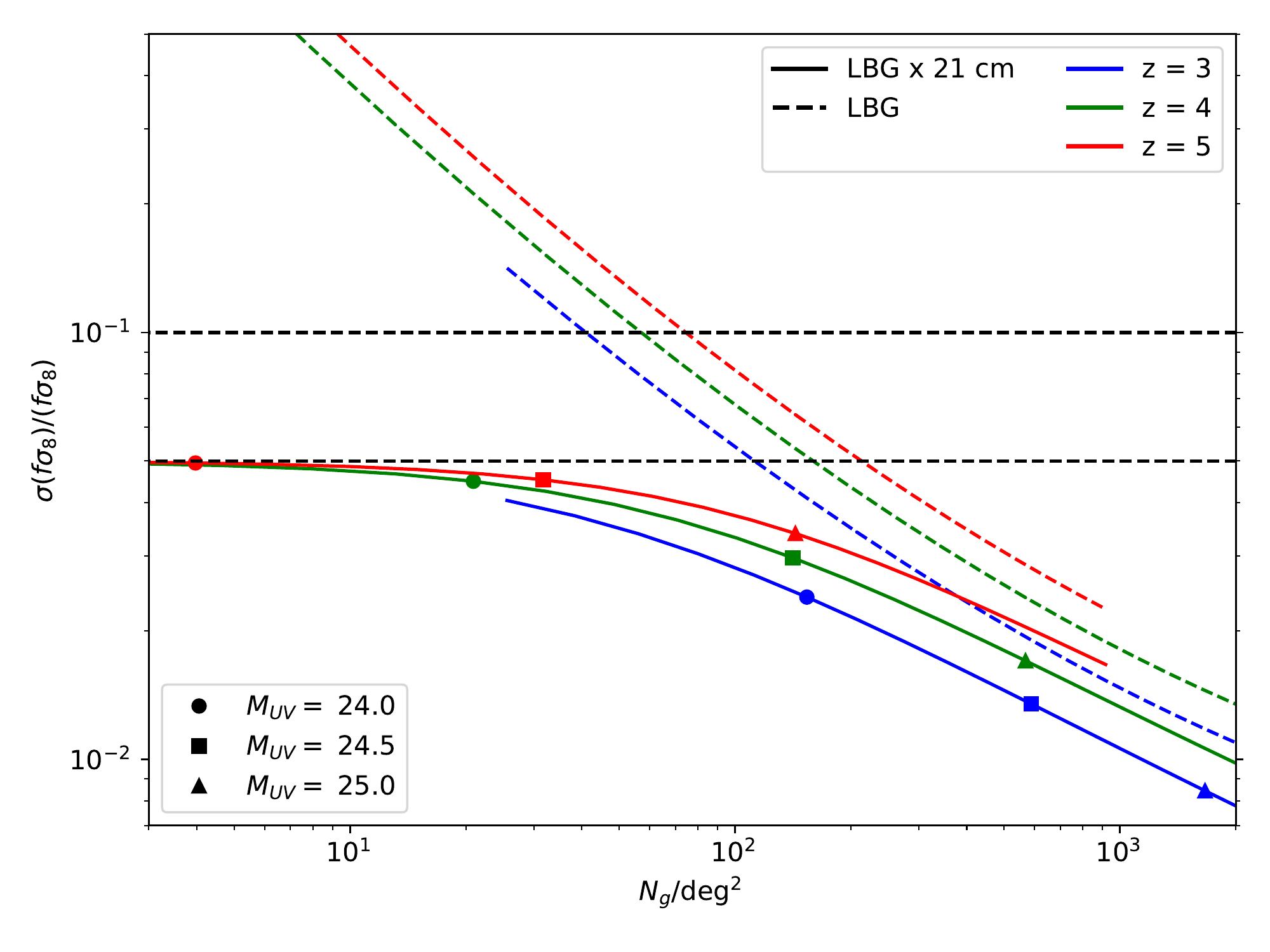}
    \caption{Constraints on the growth rate, $f\sigma_8$, at redshifts $z = 3$, 4 and 5 from a 1000 deg${}^2$ LBG survey of width $\Delta z = 0.5$, plotted against the (angular) number density of galaxies. Dashed lines indicate results from the galaxy survey alone, while solid lines show results including cross correlations with 21-cm data. Circles, squares and diamonds show the expected number densities for threshold apparent magnitudes of $m_{UV}^{\rm th} = 24.0$, 24.5 and 25.0, respectively. At high number densities, the degree to which the $\bar{T}$ degeneracy is broken is limited by the extent to which RSD can be measured using galaxies alone, leading to a close-to-constant improvement ($\approx$50\%) when 21-cm data are added at all redshifts. To a very good approximation errors scale as $f_{\rm sky}^{-1/2}$.
    }
    \label{fig:fs8_constraints}
\end{figure}

As a benchmark, we explore the merits of a 1000 square degree LBG survey at limiting magnitudes ranging between $m_{UV}^{th} = 24$ and $25.5$. The forecasted $f \sigma_8$ constraints one could obtain from such surveys, with and without additional Stage {\sc ii} 21-cm data, are plotted in Fig.~\ref{fig:fs8_constraints}.  For a wide range of sky areas the constraints scale as $f_{\rm sky}^{-1/2}$. 
At slightly more than 100 galaxies per square degree we push into sub 5\% constraints on $f \sigma_8$ at redshifts 3, 4 and 5. At $z = 3$, constraints from cross-correlating an LBG survey with $m_{UV}^{\rm th} = 24$ with Stage {\sc ii} over 1000 square degrees slightly outperforms those from cross-correlating DESI QSOs over 14,000 square degrees, and is about about 50\% better than constraints using the LBGs alone. The gains are more noticeable at $z = 4$, where including 21-cm cross correlations would bring an LBG-only constraint at 10\% to below 5\%. At the highest redshift we consider,  $z = 5$, one must push fainter than $m_{UV}^{\rm th} = 25$ to obtain sub-5\% constraints on $f \sigma_8$.  It is unclear how feasible such a survey would be.  However, even at $m_{UV}^{\rm th} = 24.5$ such a survey will be able to deliver sub-10\% constraints, more than a factor of two improvement over what can be obtained from LBGs alone, though in this regime our forecast gets close to being prior-dominated. At the highest galaxy counts shown in Fig.~\ref{fig:fs8_constraints}, the LBG-only RSD measurement itself becomes substantially better than constraints obtainable from DESI QSOs + Stage {\sc ii} data, reflecting the fact that breaking the $f\sigma_8 - \bar{T}$ constraint beyond a certain degree of accuracy necessarily involves increasingly good measurements of the amplitude of the galaxy data.  If it were possible to construct such a deep LBG survey over a wide area it would be highly competitive with Stage {\sc ii}.

\begin{figure}
    \centering
    \includegraphics[width=\textwidth]{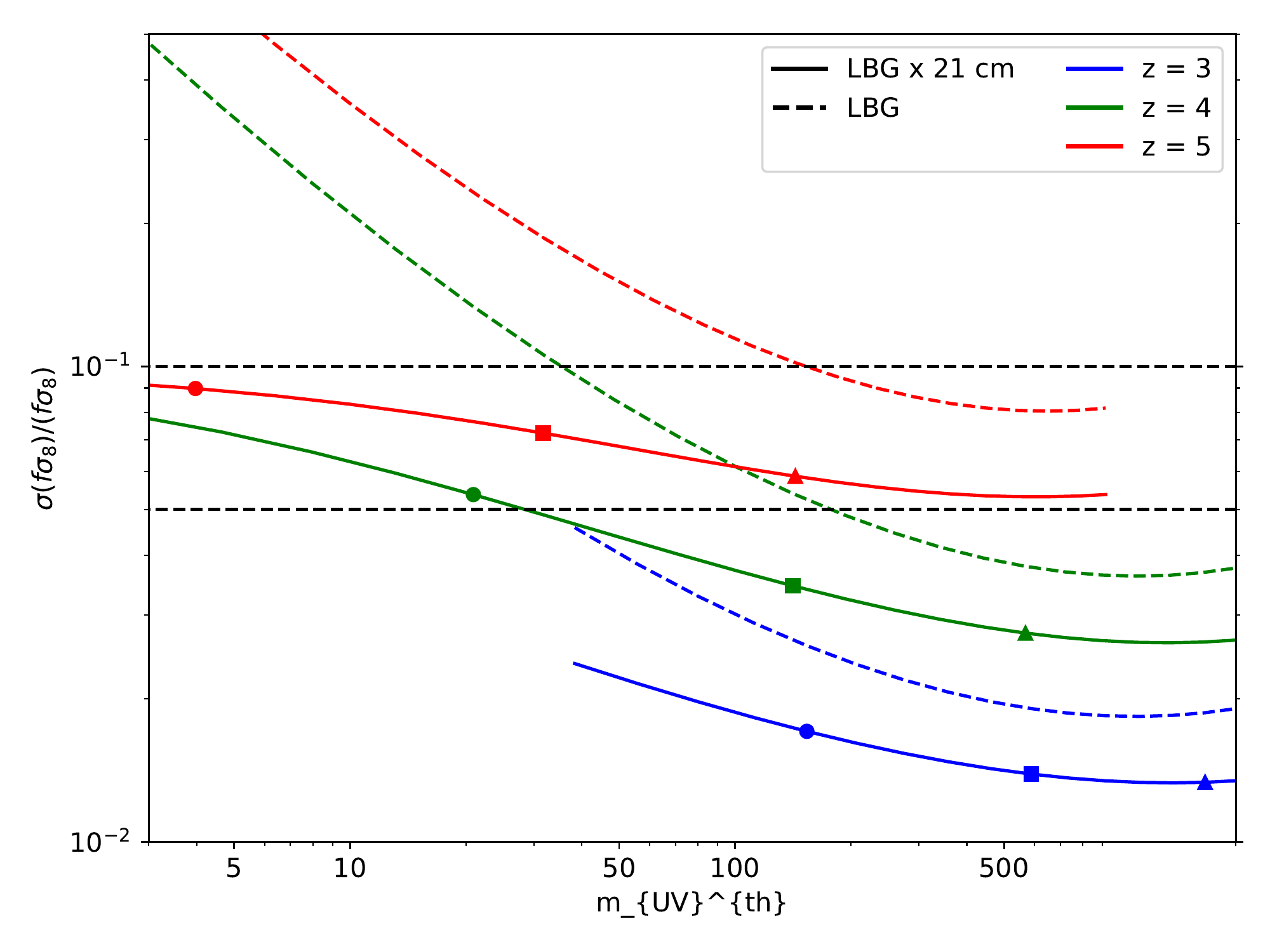}
    \caption{Same as Fig.~\ref{fig:fs8_constraints} but for fixed observing cost, i.e.~$\log_{10}f_{\rm sky}= -0.8m_{UV}^{\rm th}+$const., normalized so that 1000 square degrees are measured when the limiting magnitude is $m_{UV}^{\rm th} = 24.5$. While conducting a deeper, more limited area survey always yields better constraints at the redshifts and magnitude ranges of interest, adding 21-cm data consistently narrows constraints by $\approx 50\%$ even for the most optimally configured LBG survey at fixed cost.
    }
    \label{fig:fs8_constraints_fixedcost}
\end{figure}

We have also investigated the constraints on $f\sigma_8$ as we vary depth and area under the assumption that the LBG survey has a fixed integration time.  For simplicity we take observing time $t_{\rm obs} \propto f_{\rm sky} t_p$, where $t_p$ is the time per pointing, assuming that all galaxies within some flux limit and a given field of view are observed in each pointing. The limiting flux is assumed to be determined by the signal-to-noise that can be achieved per pointing, which is itself assumed to decrease $\propto \sqrt{t_p}$. The observable sky fraction thus scales with the limiting apparent magnitude as $10^{-0.8 m_{UV}^{\rm th}}$.
The resulting constraints are shown in Fig.~\ref{fig:fs8_constraints_fixedcost}. We have fixed the observing time at what it would take to observe down to $m_{UV} = 24.5$ over 1000 square degrees. In contrast to the fixed-$f_{\rm sky}$ analysis, the gains from conducting deeper surveys flatten out at high apparent magnitudes. At $z = 3$ there is little gain from conducting a narrower but deeper survey at $25^{\rm th}$ magnitude as opposed to a wider but shallower survey half a magnitude brighter. On the other hand, for most practical magnitude ranges it is preferable to perform deeper surveys given fixed observing time, reflecting the near exponential increase in LBG density at luminosities close to $L_{\star}(z)$. Most notably, adding cross-correlations with the Stage {\sc ii} experiment improves the forecasted RSD measurement by about 50\% in all redshift bins, even for the most optimal LBG survey configurations.
This 50\% improvement is inline with naive expectations that our ability to break the $f\sigma_8-\bar{T}$ degeneracy is set by the same factors that determine how well we can measure $f\sigma_8$ from the LBGs.

\subsection{Potential Complications}

In this section we consider two possible obstructions to measuring RSD by cross-correlating 21-cm data from the Stage {\sc ii} experiment with a spectroscopic galaxy sample. The first complication -- the extent of foreground leakage into the ``wedge'' -- comes purely from limitations of the 21-cm measurement and how well systematics from the instrument can be controlled and modeled.  The second complication arises from nonlinear RSD phenomena, particularly fingers of god in the RSD data (or redshift errors). In addition, we will discuss potential benefits of increasing the $k$-range in the fit up to the nonlinear scale.

\subsubsection{Degradation of Constraints from Foreground Leakage}

The extent to which foreground modes leak into the ``wedge'' (defined in \ref{ssec:stage_ii_exp} and shown in Fig.~\ref{fig:kplane}) will restrict the number of modes in the 21-cm data  (roughly $\propto 1 - \mu_{\rm min}$) available to constrain RSD information. In the following we will consider the effects of the foreground wedge on our $f \sigma_8$ constraints in three illustrative scenarios: the case of no foreground leakage and the ``optimistic'' and ``pessimistic'' wedges defined in \S\ref{ssec:stage_ii_exp}. To avoid confounding factors, we have chosen in this section to not include our fiducial 10\% prior on the brightness temperature $\bar{T}$ in the forecasts, which tends to put a 10\% floor on our forecasts, but will comment on its effects as is appropriate.

\begin{figure}
    \centering
    \includegraphics[width=1.0\textwidth]{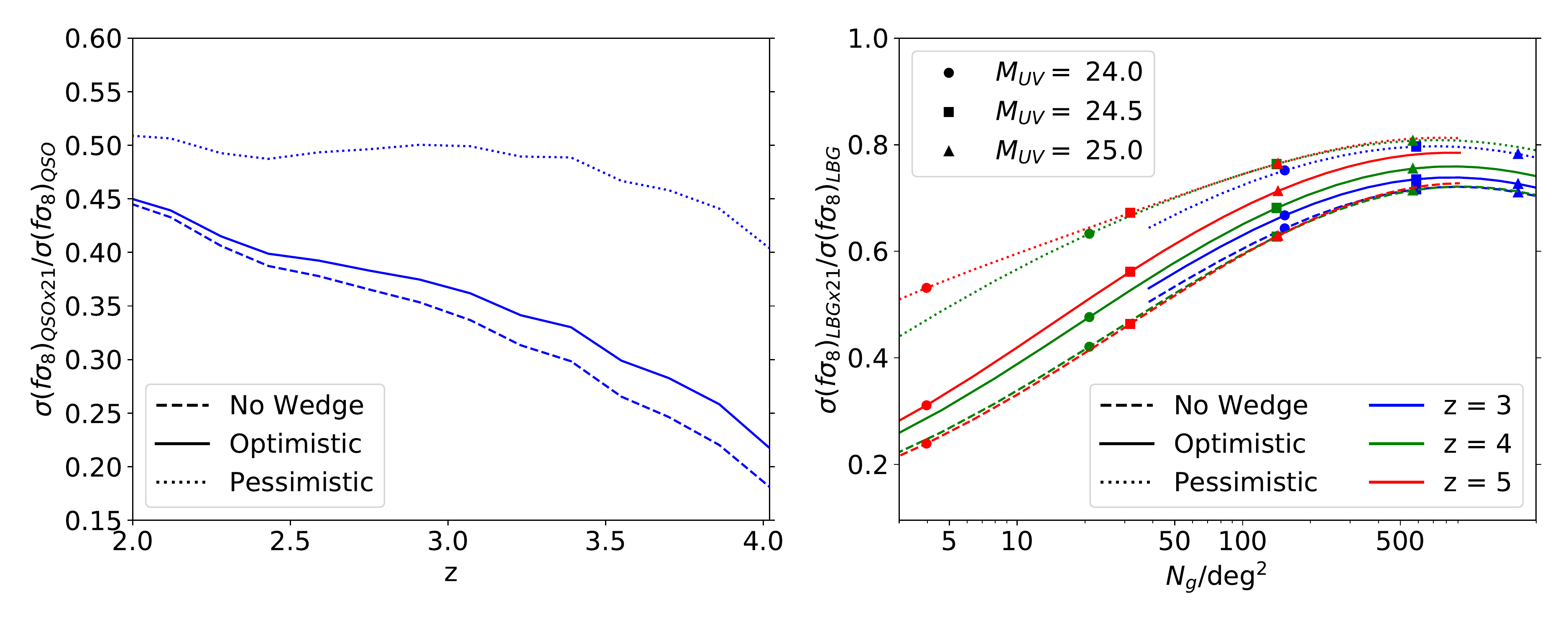}
    \caption{Ratio of constraints from galaxy$\times $21-cm data vs.~galaxies alone for three different wedge scenarios: no wedge, optimistic and pessimistic (see text; \S~\ref{ssec:stage_ii_exp}).  The left panel shows the effect of these scenarios when the galaxy sample is the expected DESI QSOs between $z=2-4$. While the optimistic wedge scenario is not significantly different from no wedge at all, the pessimistic wedge significantly hampers constraint improvements from adding 21-cm data, particularly at higher redshifts.
    The right panel shows the effect of these scenarios on hypothetical LBG surveys at $z = 3, 4, 5.$ The pessimistic wedge is similarly noticeably worse than both no wedge or the the primary wedge, which offer qualitatively similar improvements. This difference decreases at the highest number densities, however.
    }
    \label{fig:wedge}
\end{figure}

Figure~\ref{fig:wedge} shows the constraints obtainable in each of the wedge scenarios as a fraction of the $f \sigma_8$ constraint obtainable from a galaxy sample alone for both DESI QSOs and our hypothetical 1000 square degree LBG survey. As expected, the effect of these scenarios grows with redshift as the wedge takes up increasingly large fractions of $k$-space (Fig.~\ref{fig:kplane}). Intriguingly, at $z = 2$ there is next to no difference between the primary wedge and no wedge at all. More significantly, in both cases the relative degradation in the $f \sigma_8$ constraint is strongly number-density dependent -- at $z = 4$ it is close to a factor of two for the comparatively sparse DESI QSOs and at most ten to twenty percent for the LBG survey at the highest limiting magnitudes. We thus expect denser galaxy samples to be more robust against complications from foreground subtraction. A caveat, however: as the largest wedge effects coincide with where constraints are weakest, they will in practice also be the most affected by existing $\bar{T}$ priors. In this regime the $f \sigma_8 - \bar{T}$ degeneracy is primarily broken by the $\bar{T}$ prior. This has the effect of putting a lower limit on the $f \sigma_8$ constraint at the level of the prior (10\% in our case) regardless of wedge scenario. 

\subsubsection{Nonlinear Modeling and Fingers of God}

We have steered clear of nonlinear effects in our forecasts so far, both due to their uncertain nature and for simplicity of presentation. However, as the bulk of accessible $k$-modes will be at the smallest scales probed, where nonlinearities will simultaneously be most pronounced, to obtain the strongest possible constraints it is important to consider the extent to which modeling of small scales can be trusted. In this section, we will investigate both FoG effects and the potential to obtain stronger constraints by pushing to wavenumbers beyond $k = 0.4 h {\rm Mpc}^{-1}$ in spite of these effects.

\begin{figure}
    \centering
    \includegraphics[width=\textwidth]{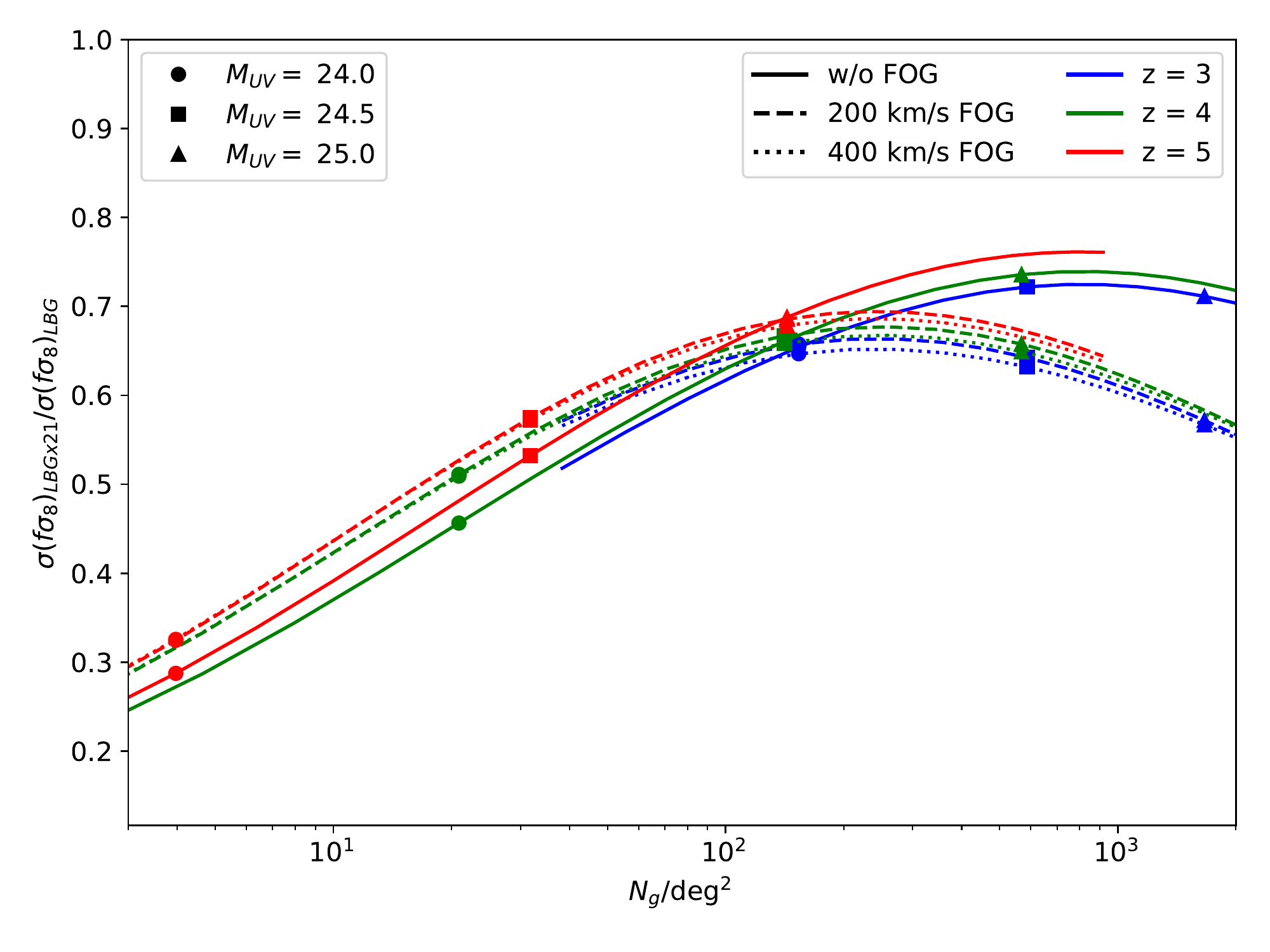}
    \caption{Ratio of constraints from LBG$\times$ 21-cm data vs.~LBGs alone for the case of no fingers of god, and for FOG with velocities at $v = 100$ and $200\,$km s$^{-1}$. The first case differs significantly from the former both at low number densities, where a degeneracy of the angular FOG suppression and the Kaiser factor are quasi-degenerate, and at the highest number densities, where constraints from LBGs alone are hindered by large FOG damping.}
    \label{fig:fog}
\end{figure}

Figure~\ref{fig:fog} compares, at fixed $k_{\rm max}$, the improvement of the $f \sigma_8$ constraint when Stage {\sc ii} data are used in the case where the LBG FOG parameter $\sigma_v = 0$ and when it is allowed to vary about values of $200\,{\rm km}\,{\rm s}^{-1}$ and $400\,{\rm km}\,{\rm s}^{-1}$. The latter two scenarios, which do not differ significantly, are qualitatively different from the scenario where FOG are set to zero.  At lower number densities, the similar angular behavior of the FOG and Kaiser factor provides a quasi-degeneracy that worsens 21-cm performance, while at higher number densities FOG damp signal power for the LBGs, yielding greater improvements when 21-cm data are available.  Importantly, due to their relative magnitude the effects discussed above do not qualitatively alter our conclusions from linear-theory forecasting.

\begin{figure}[t]
    \includegraphics[width=1.0\textwidth]{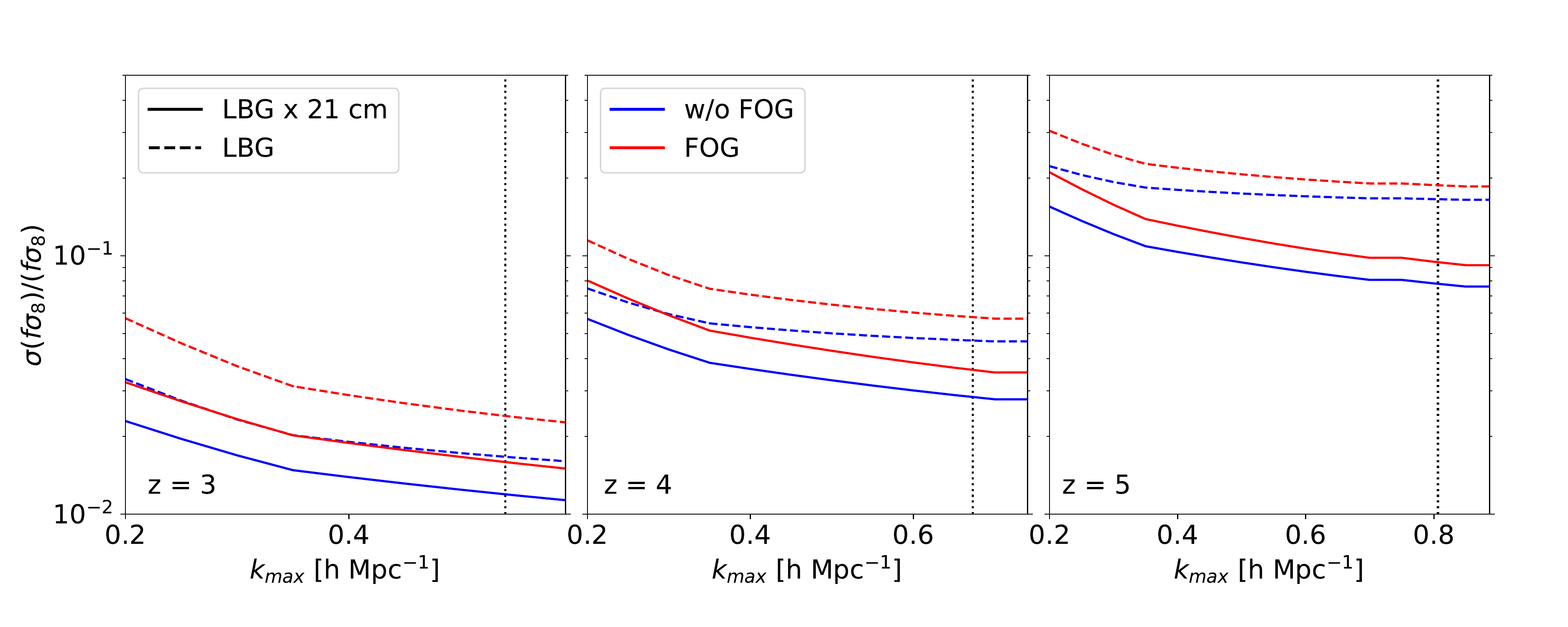}
    \caption{Constraints on $f\sigma_8$ for different values of $k_{\rm max}$ at $z = 3,4,5$ for a fiducial LBG survey at $m_{UV}^{\rm th} =24.5$ over 1000 square degrees. Black dotted lines indicate the nonlinear scale $k_{\rm nl}(z)$ at each redshift. At the highest redshifts, while constraints from LBGs alone flatten, constraints with 21-cm data included continue to improve out to the nonlinear scale, yielding a factor of two improvement between $k_{\rm max} = 0.2 - 0.4 h \; \text{Mpc}^{-1}$ at $z = 5$ when a $200$ km s$^{-1}$ FOG factor is included.
    }
    \label{fig:kmax}
\end{figure}

Figure \ref{fig:kmax} shows the degree of improvement as we increase $k_{\rm max}$ beyond our fiducial value ($0.4\,h\,{\rm Mpc}^{-1}$) for a fiducial LBG survey at $m_{UV}^{th} =24.5$ over 1000 square degrees.  We assess the improvement obtained by increasing $k_{\rm max}$ all the way to  $k_{nl} = \Sigma^{-1}$ (see Equation~\ref{eq:sigma} and the surrounding discussion). Pushing to smaller scales consistently improves constraints when 21-cm data are included up to the nonlinear scale at all redshifts, while constraints from LBGs only peter out at the higher redshifts. This is expected -- at high redshifts the LBG data become shot noise dominated well before the nonlinear scale while the Stage {\sc ii} data do not (e.g.\ Fig.~\ref{fig:snr}) -- and represents a significant advantage for using 21-cm data as a large scale structure probe in the high-redshift universe. The effect is especially stark at $z = 5$, where LBGs are sparse and the nonlinear scale is at $k_{\rm max} = 0.8 h \; \text{Mpc}^{-1}$; going from LBGs only to LBGs$\times$21-cm data improves constraints by 50\% at $k_{\rm max} = 0.2 h \; \text{Mpc}^{-1}$ but up to a factor of two at the nonlinear scale. Furthermore, while going to higher $k_{\rm max}$ decreases the gap between constraints that marginalize over FOG when only LBG data are used, the qualitative change is much less pronounced when 21-cm data are included, such that marginalizing over FOG induces constraints that are worse by more-or-less a constant factor across the explored range of $k_{\rm max}$.

\section{Synergies with the CMB}
\label{sec:cmb}

In the previous sections, we explored the extent to which cross-correlations between two biased tracers -- data from the Stage {\sc ii} 21-cm experiment and a spectroscopic galaxy survey -- could be leveraged to better measure the growth rate $f \sigma_8$ of cosmic structure. CMB lensing offers a complementary probe of the same structure with different physics -- information is primarily obtained from perpendicular as opposed to radial modes, and photon deflections are governed by the Weyl potential in contrast to clustering dynamics, which depend on the Newtonian potential alone (Eq.~\ref{eq:gamma}).

The complementary physics of redshift-space distortions and lensing provides the opportunity to break degeneracies inherent to the analysis of either on their own. In particular, the magnitude of RSD and lensing scale with the growth rate $f$ and slip parameter $\gamma$, respectively, but both also scale equally with the amplitude of spatial fluctuations, $\sigma_8$, which would thus be degenerate with both $f$ and $\gamma$ if analyzed independently. At high redshifts, $f \approx \Omega_m^{0.55}(z)$ approaches unity and, assuming $\Lambda$CDM and current measurements (e.g.~\cite{Planck18-I}) is known to better than a percent. In this section, we consider measuring $\gamma$ while holding $f$ fixed, assuming the growth of structure is governed by the Newtonian potential as in general relativity but allowing the Weyl potential to deviate.

\begin{figure}
    \centering
    \includegraphics[width=\textwidth]{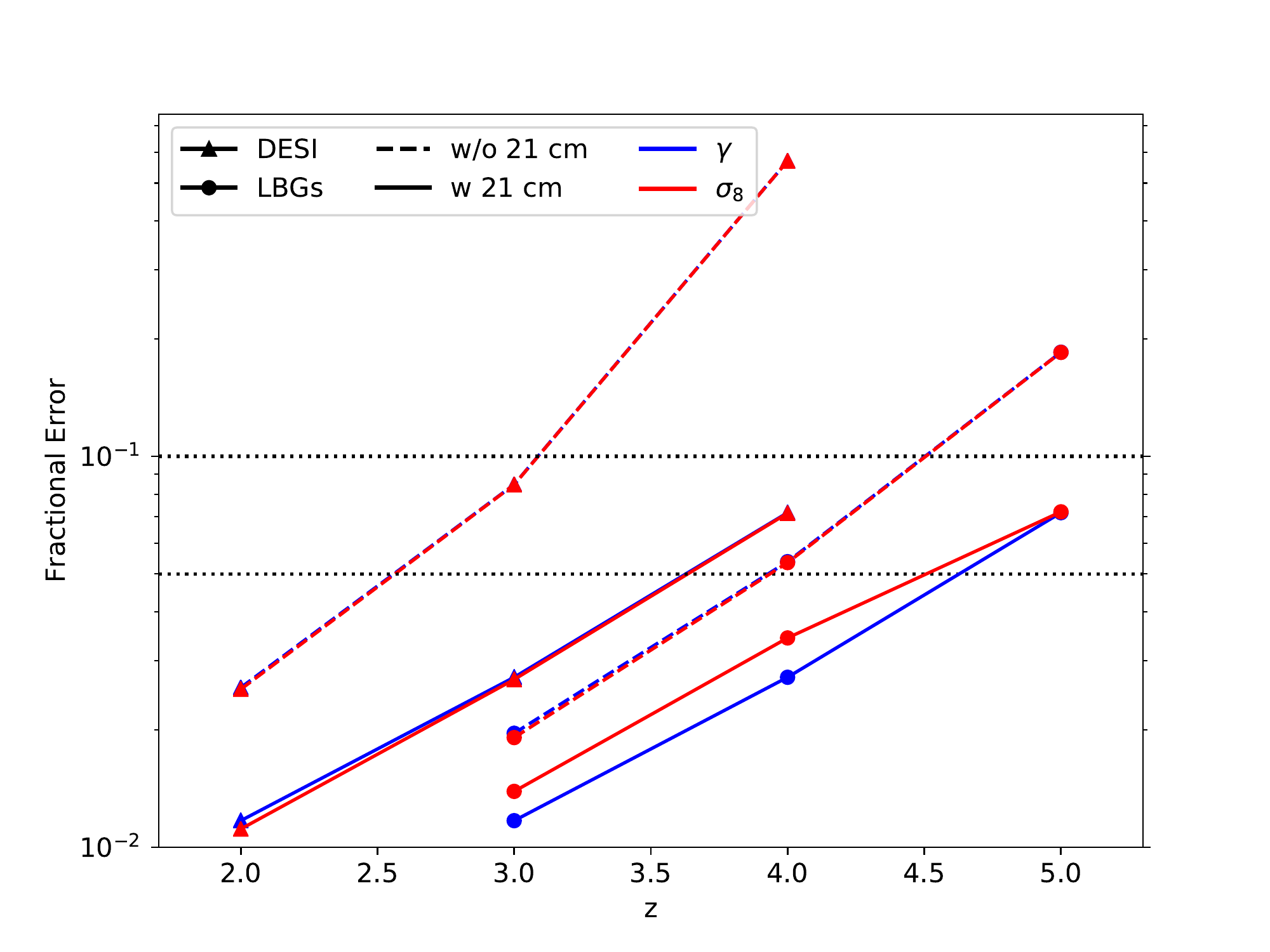}
    \caption{Forecasted constraints on the gravitational slip parameter, $\gamma$, and the power spectrum normalization, $\sigma_8$, using a combination of RSD and CMB-lensing$\times$LSST cross-correlations assuming $f(z)$ is fixed. RSD data assume DESI QSOs, a LBG survey at $m_{UV}^{\rm th} =24.5$ over 1000 square degrees, or either cross correlated with Stage {\sc ii} data. Adding 21-cm data to the RSD side significantly improves constraints on $\gamma$, yielding sub-4 percent constraints at $z=2$, 3 and $4$ from DESI QSOs and our hypothetical LBG survey at $z = 3$ and 4, respectively.  At $z=5$ adding 21-cm data improves constraints by more than a factor of two, to below ten percent.}
    \label{fig:gs8}
\end{figure}

We isolate $\gamma$ by adopting a new set of parameters $\sigma_8, \gamma, b_g, b_{\HI}, \bar{T}, b_{LSST}$ for our Fisher forecasts. Figure~\ref{fig:gs8} shows constraints on $\gamma$ and $\sigma$ using a combination of LSST-CMB lensing cross correlations, Stage {\sc ii} 21-cm and spectroscopic survey data from DESI and a representative $m_{UV}^{th} = 24.5$, 1000 square degree LBG survey. Including 21-cm data improves constraints on $\gamma$ by more than 70\% at all redshifts for both galaxy samples, and yields sub-5\% constraints at all redshifts below $z = 5$, where the constraint is at about 7\%. The fractional errors on both parameters are essentially identical and trace the $f \sigma_8$ constraint from fitting only Stage {\sc ii} and galaxy cross correlations in RSD. This can be understood roughly as follows: $\gamma \sigma_8$ roughly probes the amplitude of the lensing data, which can be constrained to sub-percent levels using CMB lensing and LSST alone\footnote{Specifically, both $b_{LSST}$ and $\sigma_8$ can be probed to sub-percent levels using this combination. See \cite{Yu18, Schmittfull18} for more details.}. The parameters can thus be individually constrained up to the level of the $\sigma_8$ constraint from RSD alone, which should be essentially equal to the $f \sigma_8$ constraint. Indeed, direct examination of the resulting Fisher matrices confirms that $\gamma$ and $\sigma_8$ are essentially degenerate.

Finally we note that it is also possible to constrain $f$ and $\sigma_8$ separately using a method identical to the one described above, but instead setting $\gamma = 1$ and allowing varying $f$ as an independent parameter. Such an analysis would, however, reveal a $\sigma_8$ constraint dominated by that from the LSST-CMB lensing cross correlations Fisher, which is significantly stronger than what can be obtained on $f\sigma_8$ by studying RSD in the Stage {\sc ii} and galaxy data. The resulting constraint on $f$ would thus essentially reduce to the level of the $f \sigma_8$ constraint from RSD alone. 

\section{Conclusions}
\label{sec:conclusions}

Interferometric surveys of neutral hydrogen can in principle map enormous volumes of the Universe in the redshift range $2<z<6$ with exquisite radial resolution and sufficient angular resolution to measure large-scale structures.  If foregrounds can be controlled and systematics well enough understood, such instruments would enable highly precise determinations of cosmological parameters and strong constraints on cosmological models.  Unfortunately the neutral hydrogen signal is modulated by a mean sky brightness, $\bar{T}$, which is only weakly constrained.  This introduces a degeneracy between $\bar{T}$ and cosmological parameters such as the growth rate, $f\sigma_8$, which needs to be broken by external data.

In this paper we have investigated the extent to which a spectroscopic survey of QSOs or LBGs, overlapping in redshift and sky area with a futuristic (Stage {\sc ii}) 21-cm survey, can break the $\bar{T}$ degeneracy and enable precise measurements of $f\sigma_8$ beyond what is possible from conventional spectroscopic surveys alone. With $N_s = 256$ six-meter dishes on each side, the Stage {\sc ii} instrument will be signal dominated up to the nonlinear scale at $z = 2-5$. In contrast, conventional spectroscopic surveys are unencumbered by $\bar{T}$-related degeneracies but become shot-noise dominated well before $k = k_{\rm nl}$, particularly at the highest redshifts we consider ($z=5$). The complementary strengths of conventional spectroscopic surveys and 21-cm experiments opens up the possibility for synergies in analyzing their cross-correlations.
 
 For DESI surveys, the improvements are a factor of 2-5 at redshift $z=2-4$, with larger gains towards higher redshifts. A realistic prior of 5\% on $\bar{T}$ helps, particularly at high redshift where it brings the improvement over DESI to an order of magnitude. Importantly, in this case one can measure $f\sigma_8$ over this redshift range with a sensitivity of better than 5\%. For the LBG case, we find more modest improvements, depending strongly the number-density of LBG sources. We find the biggest gain at high redshifts and for relatively shallow LBG surveys, where the improvements can be a factor of 5 at $z>4$ with $m_{UV}^{th}<24$. A combination of $m_{UV}^{th}<25$ LBG with a Stage {\sc ii} 21-cm experiment would result in better than 5\% measurement in $f\sigma_8$ across the full redshift range to $z=5$ without any external priors, but in this case the 21-cm data contribute significantly only in the highest redshfit bin. In an appendix, we have also investigated the extent to which this synergy can be exploited at lower redshifts using more near-term 21-cm experiments such as HIRAX or CHIME and a DESI-like spectroscopic survey in the Southern hemisphere, finding that the abundance of readily observable spectroscopic samples (i.e. ELGs) in in these epochs makes gains from additional 21-cm data more modest, a finding that will likely be exacerbated by surveys conducted through the Subaru Prime Focus Spectrograph, EUCLID or WFIRST.
 
 In addition to $\bar{T}$-related degeneracies, the analysis of 21-cm data presents a number of technical challenges, particularly with regard to the issue of foreground subtraction. We have explicitly checked the sensitivities of our forecasts to two foreground effects: the exclusion of modes at low wavenumber and within the so-called wedge. The removal of the former modes, which consist of an exceedingly small fraction of the total modes modelled, have a negligible impact. The exclusion of wedge modes, on the other hand, can have a greater impact; while forecasts taking into account the optimistic primary wedge are only marginally worse than those obtained when no wedge is taken into account, forecasts assuming a more pessimistic wedge (with $3\times$ the $\theta_w$ of the primary wedge) decreases gains by as much as a factor of two at the lowest galaxy densities considered. However, even with the pessimistic wedge adding 21-cm data still improves the forecasted $f \sigma_8$ by a factor of two, and the pessimistic wedge produces a much smaller correction when the galaxy sample is dense. In addition to these technical challenges, we have explored the extent to which the ability to model the RSD signal up to the nonlinear scale can improve our forecasts, finding that while improvements in constraints from galaxies only stall at high $k_{\rm max}$ due to shot noise, those with 21-cm data added continue to improve out to the nonlinear scale, even when nonlinear effects such as fingers-of-god in the galaxy power spectra are taken into account.
 
Finally, we also investigated how such spectroscopic surveys could serve as an intermediary between future CMB lensing measurements (which probe $k_\parallel\simeq 0$ modes) and 21-cm measurements (which probe only $k_\parallel>0$ modes).  A joint analysis all three types of surveys, observing the same sky area and redshift range, will combine the physics redshift-space clustering and lensing and enable the testing of general relativity at large scales and high redshifts, utilizing information from all sectors of $k$-space. As an illustrative example, we show that combining data from Stage {\sc ii}, a spectroscopic survey, CMB S4 lensing and LSST will let us constrain the relative amplitude of the Weyl and Newtonian potentials close to the 1\% level at $z=2$ and 3 using DESI QSOs and a $24.5^{\rm th}$ magnitude LBG survey over 1000 square degrees as the spectroscopic tracers, respectively. The latter scenario would further constrain the gravitational slip at the 3\% level at $z = 4$ and at the 7\% level at $z = 5$, sharply testing the prediction of General Relativity that masses induce spatial curvature equal to the Newtonian potential.

\section*{Acknowledgments}
EC, MW and AS would like to thank the Cosmic Visions 21cm Collaboration and all its members for providing a great and stimulating scientific environment without which this work would have not been possible, and Mike Wilson and Simon Foreman for helpful comments on an earlier draft of  this paper.
M.W.~is supported by the U.S.~Department of Energy and by NSF grant number 1713791.
SC is supported by the National Science Foundation Graduate Research Fellowship (Grant No. DGE 1106400) and by the UC Berkeley Theoretical Astrophysics Center Astronomy and Astrophysics Graduate Fellowship.
This work made extensive use of the NASA Astrophysics Data System and of the {\tt astro-ph} preprint archive at {\tt arXiv.org}.

\appendix
\section{Cross-Correlation Between Non-Overlapping Regions}
\label{sec:overlap}

Extracting the total information that can be inferred about cosmological parameters from different surveys typically involves dealing with multiple overlapping and non-overlapping regions on the sky. In this appendix, we will show that correlations between Fourier modes -- even those that sample the same underlying field when the entire sky is considered -- in non-overlapping regions on the sky can be treated as uncorrelated as a result of the mode mixing inevitably introduced by masking parts of the sky. 

Let an overdensity $\delta(x)$ be masked by some function $W(x)$ which is close to unity in the observed region and falls rapidly to zero beyond; in other words, the observed density field is $\delta^{\rm obs}(x) = W(x) \delta(x)$. If the mask is shifted by a displacement $x_0$, the observed density field is instead given by $\delta^{\rm obs}_{x_0}(x) = W(x-x_0) \delta(x)$.

In Fourier space a shift leads to a phase while multiplication by a mask becomes a convolution, such that
\begin{equation}
    \tilde{\delta}^{\rm obs}_{x_0}(k) = \int \frac{d^3k'}{(2\pi)^3} e^{-i(k-k') \cdot x_0} \widetilde{W}(k-k') \tilde{\delta}(k').
\end{equation}
The cross-correlation between two masked modes separated by a displacement $x_0$ is then
\begin{equation}
    \left\langle\tilde{\delta}^{\rm obs}_{x_0}(k) \tilde{\delta}^{\rm obs}_{0}(k')\right\rangle = \int \frac{d^3k''}{(2\pi)^3} e^{-i(k-k'')\cdot x_0} |\widetilde{W}(k-k'')|^2 P(k'') \; (2\pi)^3 \delta_D(k+k'),
\end{equation}
where the true power spectrum is defined by $   \langle\tilde{\delta}(k)\tilde{\delta}(k')\rangle = (2\pi)^3 \delta_D^{(3)}(k+k')P(k)$. For $x_0$ much larger than the width of the mask, the exponential will oscillate rapidly in $k-k''$ over the range in Fourier space where $W(k-k'')$ has support. The cross-correlation will thus be negligible and can be safely neglected.

\section{Constraints from Cross-Correlating with HIRAX}
\label{sec:HIRAX}

\begin{figure}
    \centering
    \includegraphics[width=\textwidth]{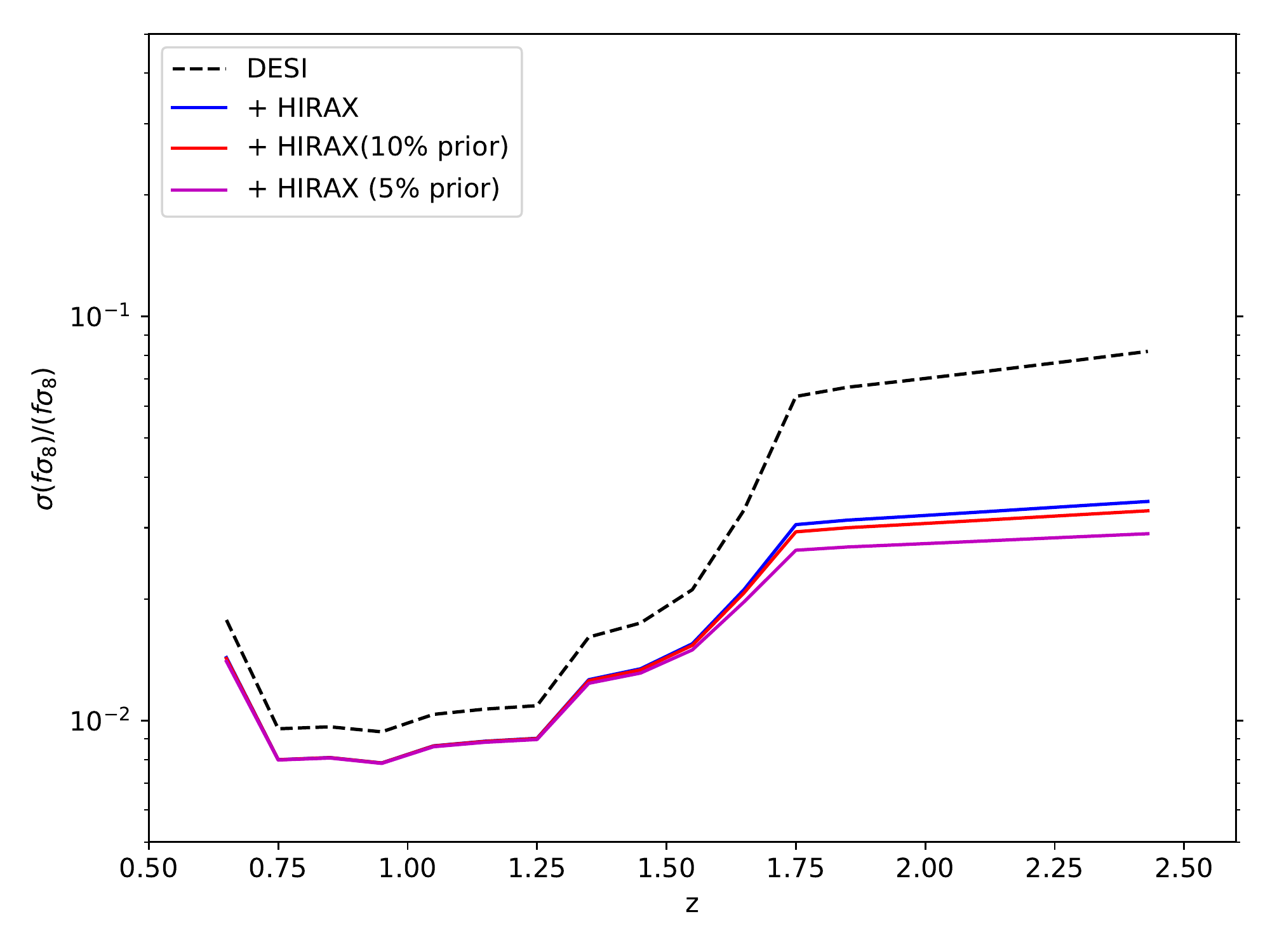}
    \caption{Constraints on $f\sigma_8$ using a DESI-like spectrscopic survey and HIRAX data, binned at $\Delta z = 0.1$, plotted for 10\% and 5\% brightness-temperature $\bar{T}$ priors, as well as no prior. While improvements from adding 21-cm data below $z = 1.6$ are at the 20\% level, above $z = 1.6$ where the observed ELG population vanishes gains from cross-correlations with HIRAX can be as great as 70\%.}
    \label{fig:hirax}
\end{figure}

In the nearer future, experiments such as CHIME \cite{CHIME} and HIRAX \cite{HIRAX} will measure the 21-cm signal between redshifts $z = 0.8-2.5$. The proposed Stage {\sc ii} experiment is in essence a scaled-up version of HIRAX, which has $N_s = 32$ six-meter dishes on each side instead of $N_s = 256$ as in Stage {\sc ii}; comparing the possible synergies of HIRAX and DESI at these lower redshifts is thus an insightful but straightforward extension of our main forecasts\footnote{We refer readers to Ref.~\cite{HIRAX} for the full noise parameters.}.  Unfortunately, there will be little expected overlap between DESI and HIRAX, which respecitvely lie in the Northern and Southern hemispheres; however, surveys such as 4MOST \cite{Roelof16} and the proposed Southern spectroscopic survey \cite{CosmicVisionsScience16} will obtain similar samples of ELGs and QSOs with large overlaps with HIRAX. For simplicity, in the following we will adopt the population and clustering statistics for DESI quoted in \cite{DESI} while assuming full overlap with HIRAX, with the assumption that the aforementioned spectroscopic surveys will be qualitatively similar. 

Both ELGs and QSOs feature prominently in the expected DESI spectroscopic sample at the redshifts probed by HIRAX, with observed ELG $dN/dz$'s dropping to zero above $z = 1.65$ while the QSO redshift distribution stays approximately constant\footnote{We take the redshift distribution for both directly from Table 2.3 of Ref.~\cite{DESI}.}. We have therefore opted to include all three distinct populations (ELGs, QSOs and the 21-cm signal) in our covariance matrices $C(k)$.  Following Ref.~\cite{DESI}, we assume $b_i(z) = b_{i,0}/D(z)$, where $b_{ELG,0} = 0.84$ and $b_{QSO,0} = 1.2$, respectively. Also, as the nonlinear scale $\propto D(z)$ is larger at low redshifts, we take $k_{\rm max} = 0.2 \; h \text{Mpc}^{-1}$ rather than $0.4\,h\;\text{Mpc}^{-1}$.

Figure \ref{fig:hirax} shows the expected constraints from DESI alone compared to those from DESI combined with HIRAX data, in bins of $\Delta z = 0.1$. At $z < 1.5$, where ELGs outnumber QSOs by more than five to one, improvements from adding HIRAX data are at the $\approx 20\%$ level. As ELGs give way to sparser QSOs, however, improvements from adding HIRAX data increase to more than 70\%, staying roughly constant past $z = 1.65$ with the DESI QSO population out to $z = 2.5.$ This confirms our intuition that 21-cm data is most helpful when the cross-correlated spectroscopic galaxies are sparse and become shot-noise dominated at lower $k$.

While we have chosen to investigate the specific example of cross-correlating HIRAX and a DESI-like spectroscopic survey in the Southern hemisphere, many other similar combinations are possible. Most directly, we expect qualitatively similar results from combining, in the Northern hemisphere, CHIME and DESI data. In the near future, the  EUCLID satellite  will similarly provide spectroscopic galaxy samples over a large sky area with which to correlate HIRAX data \cite{EuclidRed}. Finally, narrower and deeper surveys such as the Subaru Prime Focus Spectrograph (PFS) \cite{PFS14} and the WFIRST satellite \cite{WFIRST18} should likewise provide fertile ground for cross-correlations with 21-cm intensity mapping experiments in the redshift range probed by HIRAX, though the likely abundance of observed ELGs will likely imply that the largest gains will lie at high redshift..

\bibliographystyle{JHEP}
\bibliography{main}
\end{document}